\documentclass[10pt,a4paper]{article}
\usepackage[english]{babel}
\usepackage[latin1]{inputenc}
\usepackage{amsfonts,amsbsy,bm,euscript,mathrsfs}
\usepackage{amssymb,stmaryrd,faktor}
\usepackage[tbtags]{amsmath}
\usepackage{bbm}
\usepackage{graphicx}
\usepackage[title,titletoc]{appendix}
\usepackage[bookmarks=true,colorlinks=true,linkcolor=blue,citecolor=blue,urlcolor=blue,bookmarksnumbered]{hyperref}

\usepackage{dsfont}
\usepackage{collref}
\usepackage{lmodern}
\usepackage{mathrsfs}
\usepackage{mathtools}
\usepackage{bbm}
\usepackage{braket}
\usepackage{slashed}

\usepackage{graphicx}
\usepackage{booktabs}
\usepackage{subfig}
\usepackage{tikz}
\usetikzlibrary{plotmarks,calc,decorations,decorations.pathmorphing}

\numberwithin{equation}{section}

\makeatletter
\renewcommand\section{\@startsection {section}{1}{\z@}
{-3.5ex \@plus -1ex \@minus -.2ex}
{2.3ex \@plus.2ex}
{\normalfont\Large\bfseries}}
\renewcommand\subsection{\@startsection{subsection}{2}{\z@}
{-3.25ex\@plus -1ex \@minus -.2ex}
{1.5ex \@plus.2ex}
{\normalfont\large\bfseries}}
\makeatother

\newcommand{\alg}[1]{\mathfrak{#1}}

\def\e{\epsilon}

\newcommand{\beq}{\begin{equation}}
\newcommand{\eeq}{\end{equation}}
\newcommand{\beqa}{\begin{eqnarray}}
\newcommand{\eeqa}{\end{eqnarray}}

\usepackage{blkarray}

\begin{document}
%%%%%%%%%%%%%%%%%%%%%%%%%%%%%%%%%%%%%%

\thispagestyle{empty}
\begin{flushright}\footnotesize\ttfamily
DMUS-MP-22/20
\end{flushright}
\vspace{2em}

\begin{center}

{\Large\bf \vspace{0.2cm}
{\color{black} \large LonTI Lectures on Sine-Gordon and Thirring}} %v2ref
\vspace{1.5cm}

\textrm{Alessandro Torrielli\footnote{\texttt{a.torrielli@surrey.ac.uk}}}

\vspace{2em}

\vspace{1em}
\begingroup\itshape
Department of Mathematics, School of Mathematics and Physics\\University of Surrey, Guildford, GU2 7XH, UK
\par\endgroup

\end{center}

\vspace{2em}

\begin{abstract}\noindent

\end{abstract}

This is the extended write-up of a series of lectures on the duality between the Sine-Gordon model and the Thirring model. Prepared for the London Theory Institute (LonTI) - Fall 2022: a PhD-level mini-course, with exercises and a guide to the literature.

\newpage

\overfullrule=0pt
\parskip=2pt
\parindent=12pt
\headheight=0.0in \headsep=0.0in \topmargin=0.0in \oddsidemargin=0in

\vspace{-3cm}
\thispagestyle{empty}
\vspace{-1cm}

\tableofcontents

\setcounter{footnote}{0}

\section{Introduction}

We will describe the duality between two integrable systems: the two-dimensional Sine-Gordon model and the two-dimensional Thirring model. We will spend some time describing the classical and quantum Sine-Gordon model, in particular its spectrum, $S$-matrices and underlying quantum-group symmetry. We will then present the duality with the Thirring model as originally stated by Coleman and refined in subsequent literature. All the basic elements will be provided without relying on too many pre-requisites beyond standard graduate-level quantum field theory. The notes comprise a small set of embedded proposed exercises.

\section{Invitation to integrable quantum field theories}

Quantum field theory is typically not solvable in closed form and represents a very complex and extremely accurate description of elementary particle physics. Integrable quantum field theories are sufficiently simplified and yet maintain a huge richness of interesting effects. They normally reside in a $1+1$ dimensional world, and are characterised by exact $S$-matrices which are calculable in closed form starting from fundamental physical requirements such as unitarity, crossing symmetry and the location of bound-state singularities. Thanks to their vast degree of symmetry, such models are an ideal playground for training our understanding of nonpertubative quantisation. The theory of integrable systems also has a huge impact on contemporary advances in string theory \cite{Beisertetal}.

For recent reviews and textbooks on the subject the reader is invited to consult \cite{Durham,Ana,Ana2,Ana3}.

\subsection{Classical integrability}

In this section we follow \cite{AleDurham}.

Classically integrable field equations have the remarkable property that, despite being non-linear, they can be treated with exact methods and their solutions retain certain features of linear systems. This is due to the existence of an infinite number of independent classical conservation laws, encoded in an object called the transfer matrix, which in turn is built out of the so-called monodromy matrix. The monodromy matrix is constructed using the Lax pair of the system. The Lax pair is a pair of matrices $L,M$, depending on the fields, the coordinates and an additional complex variable $u$ called the {\it spectral parameter}, such that the equations of motion are equivalent to the system
\begin{eqnarray}
\partial_t L - \partial_x M = [M,L],
\label{Laxc}
\end{eqnarray} 
which singles out a flat connection.
We will see later an example of Lax pair when we will discuss the Sine-Gordon model. In general, finding such a representation of the equations of motion, if it exists at all, is almost like an art, though some constructive methods have been developed (see \cite{Babelon}). The monodromy matrix is given by the path-ordered exponential (Wilson line)
\begin{eqnarray}
T(u) = P \exp \int L \, dx.
\end{eqnarray}
We refer to \cite{AleDurham} for details on how this quantity is conserved and produces, upon a suitable expansion in $u$, a tower of conserved quantities. It is easy to see that, if we call $T_{ab} = P \exp \int_a^b L \, dx$, then, using the Lax equation (\ref{Laxc}), one gets for $\partial_t T(u)$ the result
\begin{eqnarray}
\int_a^b dx \, T_{xb} \, \partial_t L \, T_{ax}  = \int_a^b dx \, T_{xb} \, (\partial_x M + [M,L]) \, T_{ax} = \int_a^b dx \, \partial_x \big[ T_{xb} M T_{ax}\big]= M(b) T_{ab} - T_{ab} M(a). \nonumber 
\end{eqnarray}
If we set periodic boundary conditions $M(a) = M(b)$, then the result equals $[M(a),T_{ab}]$. This means that
\begin{eqnarray}
\label{virtue}
\partial_t \mbox{tr} T(u) = 0,
\end{eqnarray}
since the trace of a commutator vanishes. The dependence on $u$ has sometimes been left implicit but it should appear everywhere in the formulas above. It is now simple to see that expanding $ \mbox{tr} T(u) $ for instance in powers of $u$ produces an infinite tower of (Taylor or Laurent) coefficients\footnote{Depending on the problem, other types of expansion might be more natural.}, which are all conserved by virtue of (\ref{virtue}). 

The issue of these charges being in involution, {\it i.e.} all Poisson-commuting with one another - a requirement for Liouville integrability - is also reviewed in \cite{AleDurham}.

Let us remark that the spectral parameter is not a dynamical variable of the theory, in that it does not couple to the fields and it is completely absent from any on-shell data. It is also not uniquely defined, as one can introduce reparameterisations\footnote{The Lax pair itself is also by no means unique, partly because of a gauge freedom, as described in \cite{AleDurham}.}.  The variable $u$ can be thought of as an organising parameter for the generating function of the conserved charges. The very fact that it exists, meaning that the Lax pair permits the freedom of an arbitrary complex parameter, is probably a very deep reflection of integrability itself. Ultimately it allows for the powerful tools of complex analysis to enter the game, and it is known that integrable systems owe much of their striking features to a profound complex-analytic structure.

\subsection{Exact $S$-matrices\label{exa}}

In this section we follow \cite{Patrick,Diego}.

Quantising a classically integrable system in a way that preserves the conservation laws results in a quantum integrable system\footnote{There are of course quantum integrable systems with no classical analogue - one class of such example being spin-chains. From spin-chains one can obtain a classical integrable system by taking a continuum limit. Such classical systems then admit the original spin-chain as a (lattice) quantisation (discretisation).}. The presence of infinitely many conservation laws constrains the dynamics to a point where the scattering of quantum particles is reduced to the following:
\begin{enumerate}
\item no particle production/annihilation is admitted;
\item the initial and final sets of momenta are preserved - momenta are only reshuffled;
\item {\it factorisation:} the $N$-body $S$-matrix factorises in a sequence of $2$-body processes - the ordering of the sequence does not matter by virtue of the Yang-Baxter equation (see later).
\end{enumerate}
It is important to remark that these properties refer to the {\it full} (exact) scattering amplitude between particles of the {\it exact} quantum spectrum, while individual Feynman graphs do of course display production/annihilation of perturbative excitations - since the Lagrangian description has typically got interaction vertices. The resummation of the perturbative series however is bound by symmetries to respect the above constraints.
 
All we need to study is therefore the $2$-body $S$-matrix for all the particles in the spectrum. Since there are internal degrees of freedom (spin, flavour, {\it etc.}), the $S$-matrix is a matrix whose entries depend on the momenta of the scattering particles. Such matrix, which acts on the tensor product of the two vectors spaces $V_1 \otimes V_2$ representing the internal degrees of freedom of each scattering particle $1$ and $2$, can still be very complicated.  
  
If we focus on relativistic theories, we can parametrise the energy and momentum of a particle in $1+1$ dimensions as $E_i=m_i \cosh \theta_i$ and $p_i = m_i \sinh \theta_i$, where $\theta_i \in \mathbbmss{R}$ is the {\it rapidity} and $i=1,2$ labels the particle. Notice that we have explicitly assumed that {\it \underline{the masses are non-zero}} - otherwise an entire new story will have to apply (see much later in these notes). Since a Lorentz boost shifts the rapidity by a constant, relativistic invariance then dictates that the $S$-matrix depends only on the difference of the rapidities of the scattering particles: 
\begin{eqnarray}
S_{12} = S_{12}(\theta_1 - \theta_2) = S_{12}(\theta), \qquad \theta = \theta_1 - \theta_2. 
\end{eqnarray} 
\begin{figure}[h]
    \centerline{\includegraphics[width=6cm]{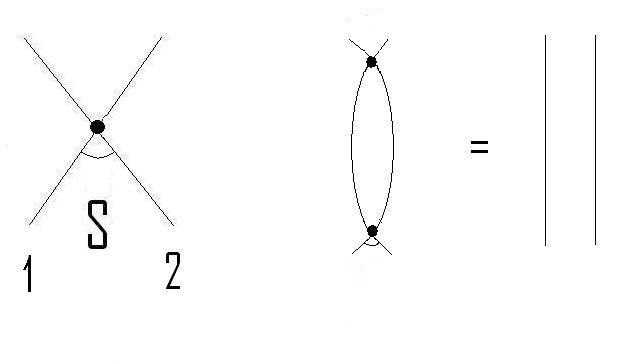}}
    \caption{The $2$-body $S$-matrix and its braiding unitarity property.}
    \label{fig3}
\end{figure}
\begin{figure}[h]
    \centerline{\includegraphics[width=6cm]{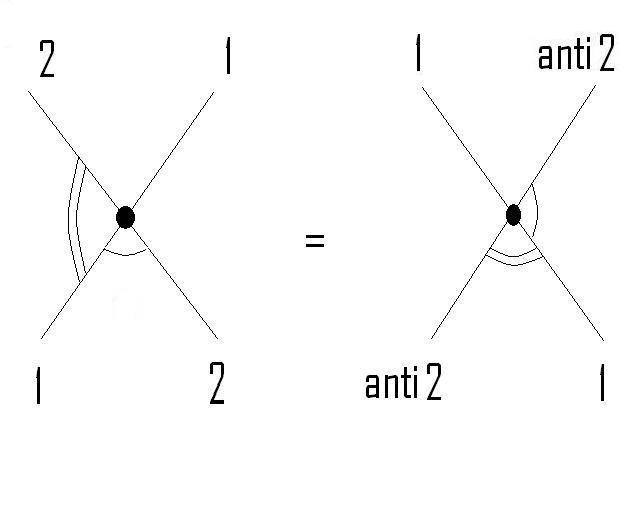}}
    \caption{Crossing symmetry: the image on the left is in the direct ($s$-)channel, while the image on the right is seen from the perspective of the crossed ($t$-)channel (rotating by $90$ degrees).}
    \label{fig4}
\end{figure}
\begin{figure}[h]
    \centerline{\includegraphics[width=6cm]{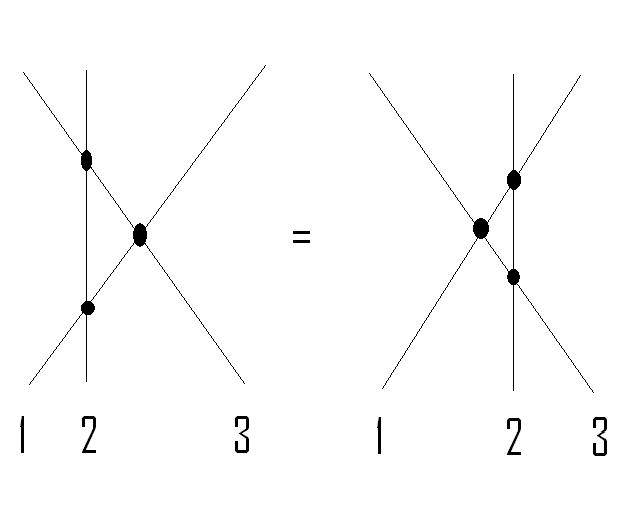}}
    \caption{The Yang-Baxter equation, which says that the ordering of the sequence of $2$-body scatterings does not matter.}
    \label{fig5}
\end{figure}
\begin{figure}[h]
    \centerline{\includegraphics[width=6cm]{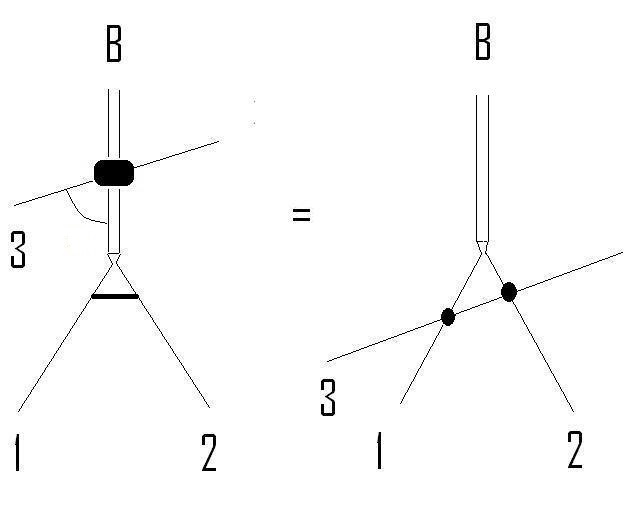}}
    \caption{The bootstrap principle: the (hyperbolic) angle denoted by a thick line is the value at which ${\bf B}$ appears as a simple pole in $S_{12}$. All angles of the image are then fixed once the scattering angle of $3$ with ${\bf B}$ is specified.}
    \label{fig6}
\end{figure}
The ``axioms'' (the word being taken with caution, as we will later comment) of integrable scattering allow one to {\it construct} the $S$-matrix from first principles, knowing the symmetries and the particle content of the model. Such axioms are \underline{schematically}\footnote{The language of Hopf algebras and their representations, which we will see later on, allows for a mathematical translation of these axioms in a way which can be implemented using standard representation-theoretical tools.} given by
\begin{enumerate}
\item braiding unitarity: $S_{12}(\theta)S_{21}(-\theta) = \mathbbmss{1} \otimes \mathbbmss{1}$, where $\mathbbmss{1}$ is the identity matrix in the internal space of each particle - see fig. \ref{fig3};
\item crossing symmetry: $S_{12}(\theta) = S_{\bar{2}1}(i \pi - \theta)$, where $\bar{2}$ is the antiparticle of $2$ - see fig. \ref{fig4};
\item Yang-Baxter equation: $S_{12}(\theta_1 - \theta_2)S_{13}(\theta_1 - \theta_3)S_{23}(\theta_2 - \theta_3) = S_{23}(\theta_2 - \theta_3)S_{13}(\theta_1 - \theta_3)S_{12}(\theta_1 - \theta_2)$, see fig. \ref{fig5};
\item physical unitarity: the $S$-matrix is a unitary matrix for real values of the rapidities; 
\item bootstrap: if the particles $1$ and $2$ can form a bound state ${\bf B}$, the $S$-matrix of particle $3$ scattering against ${\bf B}$ is given by  $S_{3{\bf B}}(\theta) = S_{32}(\theta + i x)S_{31}(\theta + i y)$, where $x$ and $y$ are calculated knowing the (complex) rapidity difference $\theta_1-\theta_2$ at which the $S$-matrix $S_{12}$ has the bound state pole - see fig. \ref{fig6}. The reader is also invited to consult \cite{Karowski:1978ps};
\item as always in scattering theory it is useful to complexify all the momenta and exploit the power of analyticity; the $S$-matrix is a meromorphic (matrix-valued) function in the complex $\theta$-plane, with possible poles and zeros. Bound states correspond to simple poles in specific (spin, colour, flavour) channels on the imaginary segment Im $\theta \in (0,\pi)$, Re $\theta = 0$; the analytic structure in the Mandelstam plane does also have cuts, and is depicted in fig. \ref{figurante}. The entire visible portion of the Mandelstam plane is mapped to the so-called {\it physical strip} $\mbox{Im} \, \theta \in (0,\pi)$ in the complex $\theta$-plane.
\end{enumerate}

\smallskip

\begin{figure}[h]
    \centerline{\includegraphics[width=6cm]{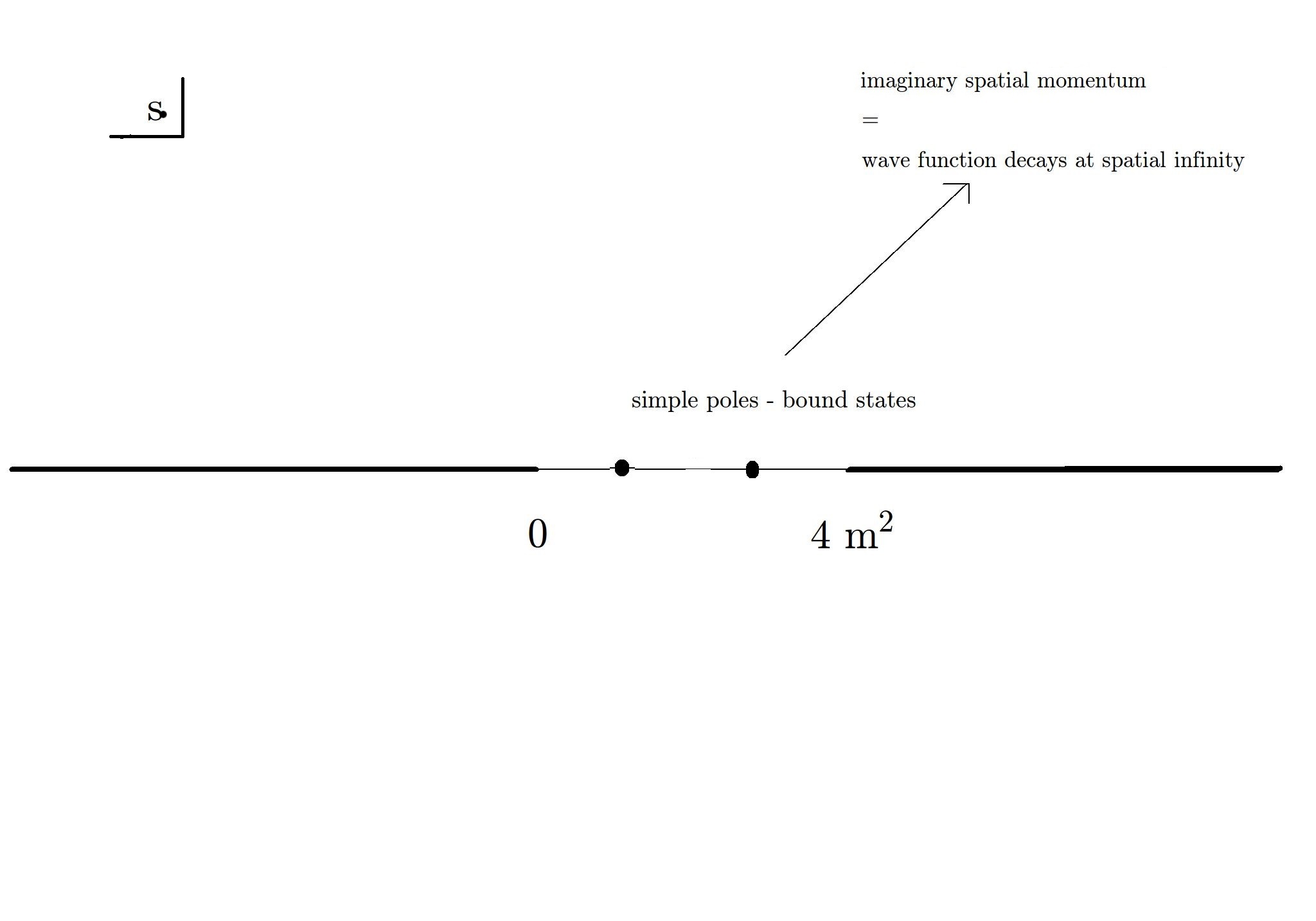}}
    \caption{The analytic structure in the Mandelstam variables s (center of mass energy). If the mass the two scattering particles is the same, then s $= 2m^2(1+\cosh \theta)$. The branch point starting at $4m^2$ (s-channel cut) corresponds to the threshold for two particles going on-shell, while the cut starting from $0$ is the correponding t-channel cut, present because of crossing symmetry.}
    \label{figurante}
\end{figure}

Notice that in two-dimensional integrable scattering (with same mass) one has (since $p_3=p_1$ and $p_2=p_4$, and given that we use the signature with $-1$ along the time direction, such that $p_i^2 = - m^2$)
\begin{eqnarray}
s = -(p_1+p_2)^2, \qquad t =-(p_1-p_4)^2 = -(p_1-p_2)^2 = 4m^2 -s, \qquad u = -(p_1-p_3)^2=0. 
\end{eqnarray}

\begin{figure}
\centerline{\includegraphics[width=17cm]{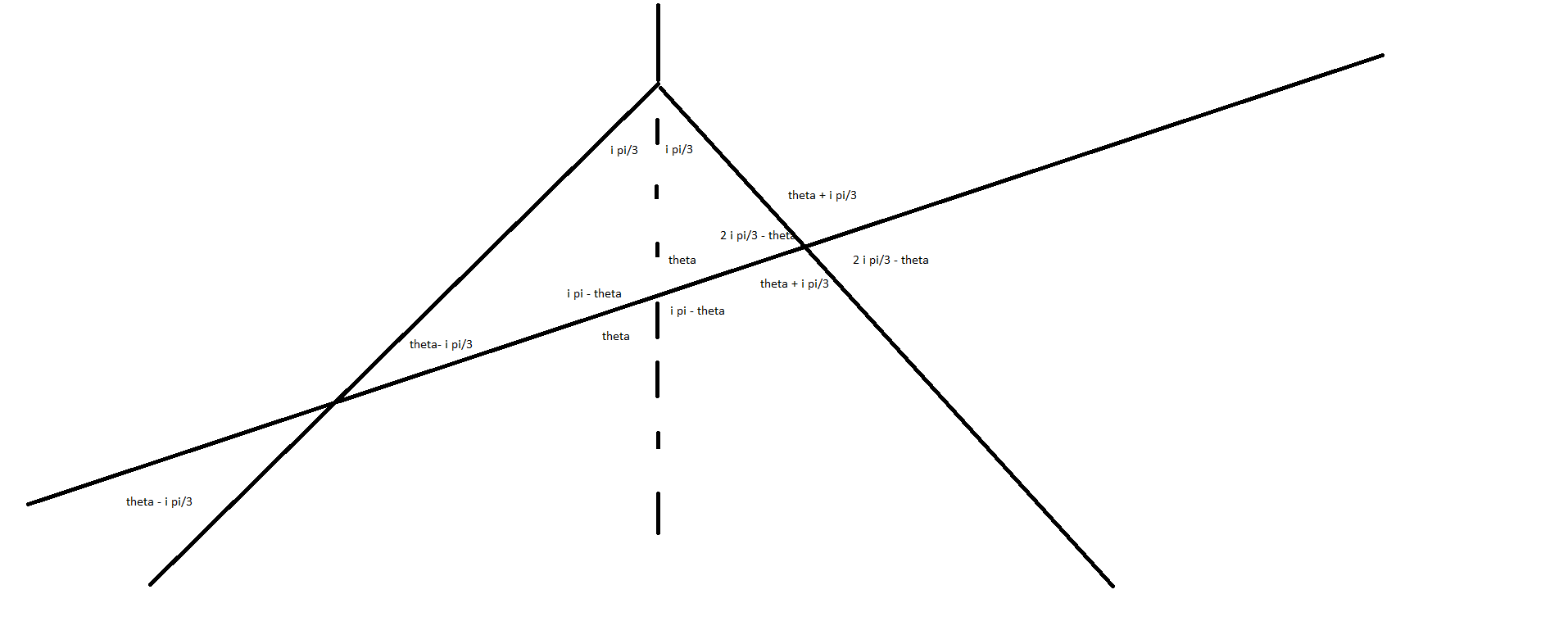}}
\caption{The bootstrap principle (see the diagram after the equal sign in fig. \ref{fig6}) for the Lee-Yang model is effectively based on simple ``angular'' relations.}
\label{figurabootstrap}
\end{figure}

To give an example of bootstrap, one can take the $S$-matrix of the Lee-Yang model, which involves a single scalar particle (no matrix structure) which is a bound state of itself:
\begin{eqnarray}
S_{LY}(\theta) = \frac{\sinh \theta + i \sin \frac{\pi}{3}}{\sinh \theta - i \sin \frac{\pi}{3}}, \qquad S_{LY}(\theta) = S_{LY}\Big(\theta+i\frac{\pi}{3}\Big)S_{LY}\Big(\theta-i\frac{\pi}{3}\Big),\label{inbo}
\end{eqnarray}
where the simple pole at $\theta = \frac{2}{3} i \pi$ is the bound state, which is the very same particle as $1$ and $2$. The bootstrap condition in (\ref{inbo}) has an easily workable graphical construction - see fig. \ref{figurabootstrap}. For a thorough discussion of the integrable $S$-matrix for a particle which is a bound state of itself the reader is invited to consult chapter 18 in \cite{Mussardo}. 

{\it Guided Exercise [3 hour's work]: Following section 2 of \cite{Pedro}, prove that the residue of $S_{LY}$ at the bound state pole has the ``wrong'' sign. This signals that unitarity ultimately breaks down - the way in which it does so being rather subtle, as explained in chapter 18 of \cite{Mussardo}.}

At this stage we can provide a justification as to why we have restricted ourselves to $1+1$ dimensions. Quantum integrable systems are characterised by a vast number of charges which commute with the quantum Hamiltonian and amongst themselves. In higher dimensions, such as $3+1$ for instance, the presence of these charges would typically imply a trival scattering matrix via the Coleman-Mandula theorem \cite{ColemanMandula}. In $1+1$ dimensions such theorem is avoided (even without invoking supersymmetry). Very roughly, the reason is that in one space and one time dimension it is quite hard to deflect two particle trajectories away from each other - see \cite{Patrick,Diego}. No matter how many charges (symmetry transformations) one might apply, the two particles will almost invariably meet. What the conservation laws will produce is a great number of constraints on the functional form of the (non-trivial) scattering matrix, most of which constraints are then subsumed into the scattering ``axioms''. 

Let us also mention that the axioms reported above are a somewhat reductive list. They can be extended to non-relativistic models \cite{Foundations} and ultimately they can be formalised into the language of Hopf algebras \cite{Diego,Loebbert,Fabian} - see also section \ref{qgroup}. Hopf algebras no longer essentially distinguish whether one is dealing with relativistic field theories, or quantum-mechanical spin-chains, or Galileian models. A comment is however necessary at this stage.  Even remaining within the domain of relativistic models, the axioms as we have presented them here are ultimately not to be considered with mathematical rigidity. There are models where not all the ``axioms'' are satisfied, yet such models may still describe interesting physics. While some axioms must \underline{always} be there (such as the Yang-Baxter equation, or the principles of local relativistic field theory such as physical unitarity and crossing symmetry), others might be relaxed if the physics so requires - these models might for instance be described by mathematical constructions even more general than Hopf algebras. Likewise, it is hard to establish any sort of strict mathematical implication of the sort ``integrable (existence of infinitely many quantum symmetries) $\leftrightarrow$ axioms''. This is because one can never exclude very singular models, which might evade the mathematical implication for some peculiar reason.  The spirit in which the construction should be taken is probably best described as follows: given any specific theory, a stubborn go at perturbation theory would ultimately reaveal which properties are verified and which are not. The ``axioms'' shortcut the whole process, by asking us to take a little (and rather well-informed) leap of faith, and allow us to boldly derive exact expressions (to later be perturbatively tested to our heart's content).     

This $S$-matrix programme (which at one point in history held the promise to become the dominant paradigm for particle physics \cite{Polkinghorne}) finds a life of its own in two-dimensional physics. The idea is therefore to derive the $S$-matrix exactly from symmetries (we will say more on this  in section \ref{qgroup}) and to impose the above ``axioms''. Parameters which are left undetermined are then matched with perturbation theory or semiclassics. One important ambiguity left over by this approach is the one represented by the so-called Castillejo-Dalitz-Dyson (CDD) \cite{CDD} factors: these are scalar factors by which any $S$-matrix can be multiplied while still preserving its basic properties. They can be seen to take the general form
\begin{eqnarray}
S_{CDD}(\theta)=\prod_{i=1}^M \frac{\sinh \theta + i \sin \alpha_i}{\sinh \theta - i \sin \alpha_i}, \qquad \alpha_i \in (0,\pi) \, \, \forall \, \, i=1,...,M, \qquad M \in \mathbbmss{N},\label{cDd}
\end{eqnarray}
in a relativistic-invariant theory. One can easily see that they are pure phases for real rapidities, and
\begin{eqnarray}
S_{CDD}(-\theta) = S^{-1}_{CDD}(\theta), \qquad S_{CDD}(i\pi - \theta) = S(\theta),
\end{eqnarray}
hence they neither spoil physical nor braiding unitarity, nor crossing.
What changes however is the pole structure, hence the physics is in principle dramatically altered. Needless to say these factors can be used to adjust a proposed solution in such a way that it reflects the expected spectrum of bound states (argued for instance using a semi-classical or a perturbative reasoning). 

\section{The Sine-Gordon model}

The Sine-Gordon model is certainly one of the most famous integrable field theories of all. Let us begin with a very nice intuitive argument from \cite{Patrick} as to why this particular model should be integrable. If one starts with the familiar scalar theory in $1+1$ dimensions 
\begin{eqnarray}
-\frac{1}{2} \partial_\mu \phi \partial^\mu \phi - \frac{m^2}{2} \phi^2 - \frac{\lambda}{4!} \phi^4, 
\end{eqnarray}  
one can compute the tree-level production amplitude $2 \to 4$. Such amplitude turns out to be constant, and it can be cancelled by adding a term $- \frac{\lambda^2}{6! m^2} \phi^6$ to the original Lagrangian. At this point one looks at the $2 \to 6$ amplitude in this new theory, and discovers that this amplitude too can be cancelled by adding a specific $\phi^8$ term. The procedure can continue until one has by hand eliminated all the possible tree-level production amplitudes, of course at the cost of having added an infinite number of additional terms to the original Lagrangian. Those infinitely many terms form the Taylor series of $-\frac{m^2}{\beta^2}(\cosh \beta \phi - 1)$, where $\beta^2 = \frac{\lambda}{m^2}$. The Sine-Gordon theory is then recovered by continuing $\beta \to i \beta$. In a sense, the infinitely many symmetries of integrability tie all the interaction monomials into being the Taylor series of a very specific function. Along the same lines one sees that quantum integrability is a symmetry that binds the renormalisation of all the monomials to proceed in unison.     

Page 4 in \cite{Patrick} explains why the argument, which we have just sketched above, does {\it not} imply the vanishing of the $3 \to 3$ amplitude (which is in fact quite importantly non-zero and the whole basis for writing the Yang-Baxter equation). The superficial expectation that the $3 \to 3$ amplitude be obtained by analytically continuing the $2 \to 4$ amplitude fails in two dimensions in integrable kinematics, due to a subtle resolution of singularities in the Feynman diagrams \cite{Patrick}.  

These arguments of cancellation of amplitudes do extend to loop level as well, but it is easier to resort to other than diagrammatic methods \cite{Fadde,Charges}. The powerful tools of the Lax matrix and the quantum inverse scattering method can then be employed. We shall work throughout these notes in particular conventions which should align, as much as and wherever possible, with Coleman \cite{Coleman}. Ultimately the Thermodynamic Bethe ansatz \cite{Zamo} furnishes the complete description of the spectrum of the model compactified on a spatial circle \cite{Ravanini}. 

\subsection{Classical aspects}

The classical Sine-Gordon Lagrangian is given by 
\begin{eqnarray}
L_{SG} = -\frac{1}{2} \partial_\mu \phi \partial^\mu \phi + \frac{m^2}{\beta^2} (\cos \beta \phi - 1), \qquad m,\beta \in \mathbbmss{R}, 
\end{eqnarray}
where $\phi = \phi(x,t)$ is a scalar field in (flat) $1+1$ dimensions. We remind that we work in the signature with time having the minus sign.
We can assume $\beta>0$ since really only $\beta^2$ matters to the Lagrangian. The $-1$ shift sets to zero the energy of the trivial vacuum $\phi=0$. Both the field $\phi$ and the parameter $\beta$ have engineering mass-dimension $0$, while $m$ has engineering mass-dimension $1$.

The equations of motion (giving the name to the model) are
\begin{eqnarray}
\partial_t^2 \phi - \partial_x^2 \phi + \frac{m^2}{\beta} \sin \beta \phi = 0
\end{eqnarray}
and tend to those of a free massive Klein-Gordon field of mass $m$ as $\beta \to 0$. The classical Lagrangian expands as
\begin{eqnarray}
L_{SG} = -\frac{1}{2} \partial_\mu \phi \partial^\mu \phi - \frac{1}{2}m^2 \phi^2 + \frac{1}{24} m^2 \beta^2 \phi^4 + ...
\end{eqnarray}
Truncating at the order $\phi^4$ now clearly does not produce a potential which is bounded below, but we do not worry about this, since we will be interested in the whole (integrable) potential which must include all the higher order corrections (and is clearly bounded). 

\subsubsection{Solitons}

\begin{figure}
\centerline{\includegraphics[width=8cm]{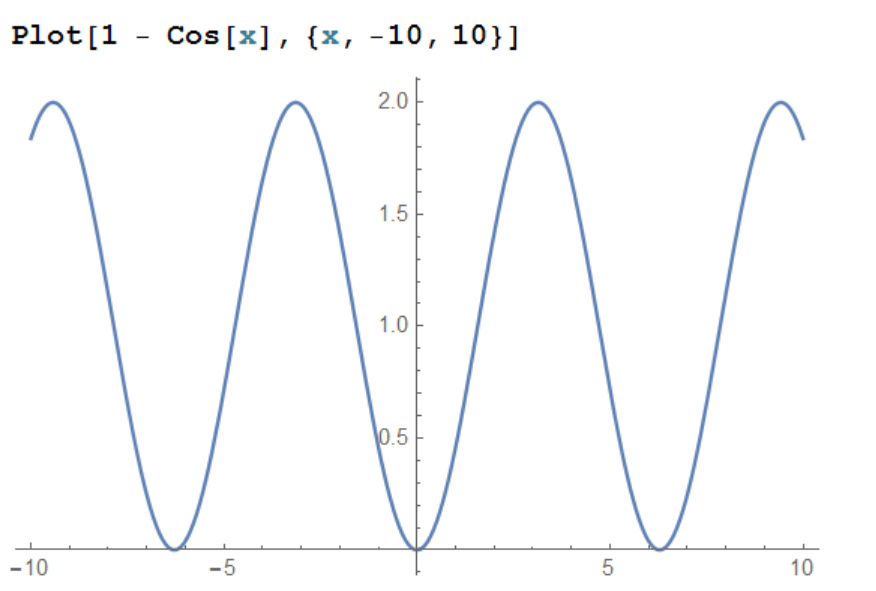}}
\caption{The Sine-Gordon potential in suitable units}
\label{fig1}
\end{figure}

\begin{figure}
\includegraphics[width=7cm]{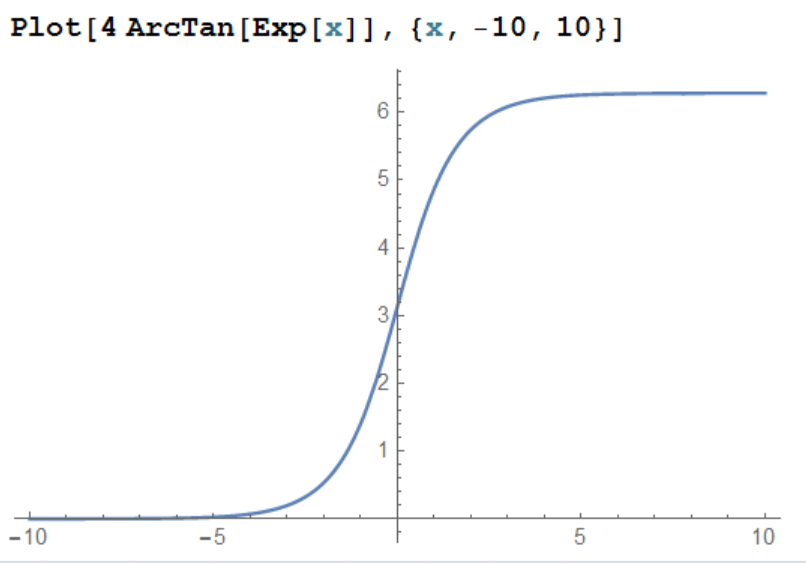}\hfill{\includegraphics[width=7cm]{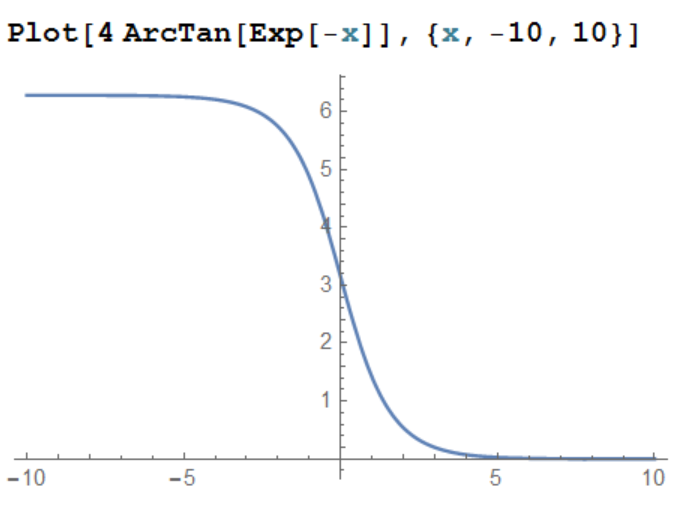}}
\caption{The Sine-Gordon static kink and, respectively, antikink, for $a=0$ and in suitable units.}
\label{fig2}
\end{figure}

The solutions corresponding to solitons and antisolitons are obtained imposing the asymptotic conditions 
\begin{eqnarray}
\label{asym}
\phi \to 0 \quad x \to -\infty, \qquad \phi \to \frac{2\pi}{\beta} \quad x \to \infty,
\end{eqnarray}
or
\begin{eqnarray}
\label{asym2}
\phi \to \frac{2\pi}{\beta} \quad x \to -\infty, \qquad \phi \to 0 \quad x \to \infty, \qquad
\end{eqnarray}
for all $t$, where (\ref{asym}) is for the soliton and (\ref{asym2}) for the antisoliton. There is a {\it topological} charge associated with these two solutions, given by 
\begin{eqnarray}
\label{charge}
q = \frac{\beta}{2\pi} \int_{-\infty}^\infty dx \, \phi' = \frac{\beta}{2\pi} \big[\phi(x\to\infty)-\phi(x\to -\infty)\big] = \pm 1,
\end{eqnarray}
where we have used the asymptotic conditions. Because of the shape of the profile these solutions are also called {\it kink} and {\it antikink}, respectively. Kink solutions exist because the classical potential has distinct minima - see fig. \ref{fig1}, and the kink/antikink interpolate between the minimum at $\phi = 0$ and the two adjacent minima:
\begin{eqnarray}
V(\phi) = \frac{m^2}{\beta^2} (1 - \cos \beta \phi) \qquad \longrightarrow \qquad \phi_{min} = 2 n \frac{\pi}{\beta}, \quad n\in \mathbbmss{Z}.
\end{eqnarray}
One can verify by brute force for instance that (for $m=1$) the two static profiles
\begin{eqnarray}
\phi = \frac{4}{\beta} \mbox{arctan} \, e^{\pm(x-a)}, \qquad a \in \mathbbmss{R},
\end{eqnarray}
satisfy the equations of motion with the asymptotes (\ref{asym}) and (\ref{asym2}) respectively - see fig. \ref{fig2}. The parameter $a$ is the centre of the kink.

Notice that, since the Lagrangian is invariant under $\phi \to -\phi$, solutions which are not invariant under this map come in pairs $(\phi,-\phi)$. One can check that
\begin{eqnarray}
-\phi_{kink} = \phi_{antikink} - \frac{2\pi}{\beta},
\end{eqnarray}  
hence $-\phi_{kink}$ is also a solution (equal to an antikink pushed down by an allowed shift-symmetry of the action). This is very clearly visible from the plots in fig. \ref{fig2}. Notice also that, the equation of motion being non-linear, we cannot make linear combinations which are eigenstates of $\phi \to - \phi$. Nevertheless, integrability still allows a certain notion of ``superimposing'' (anti)kink solutions (see later). 

A non-zero velocity can be given to the solutions \cite{DashenHasslacherNeveu}:
\begin{eqnarray}
\phi = \frac{4m}{\beta} \mbox{arctan} \exp \Big(\pm m \frac{x - a - v t}{\sqrt{1-v^2}}\Big), \qquad a \in \mathbbmss{R}, \qquad v \in (-1,1).
\end{eqnarray}
We can see the typical feature of integrable solitons to rigidly translate with their shape unmodified. Notice also a very interesting feature of the solution: it has a particle-like behaviour \cite{DashenHasslacherNeveu}.

The charge $q$ (\ref{charge}) corresponds to the current $j^\mu = \epsilon^{\mu \nu} \partial_\nu \phi$, which is conserved by virtue of the commutativity of partial derivatives. Normally such a charge would vanish for normalisable solutions, but in this case it acquires a topological meaning - see pages 18-19 of \cite{Manton}. 

The attribute of {\it topological} for the charge $q$ stems from the fact that the field $\beta \phi$ can effectively be thought of as a (target space) angle. The soliton and antisoliton profiles are contained within the fundamental region $[0,2\pi]$, and describe this angle winding the $S^1$ target space once in the positive, resp. negative direction, as $x$ goes from $-\infty$ to $\infty$. The conserved charge $q$ is just the winding number of the field regarded at any fixed time as a map $\beta \phi: \mathbbmss{R}^{1,1} \to S^1$. The conservation of this number and its additivity as a charge acquire therefore a topological nature \cite{Manton}.
 
\begin{figure}
\includegraphics[width=5cm]{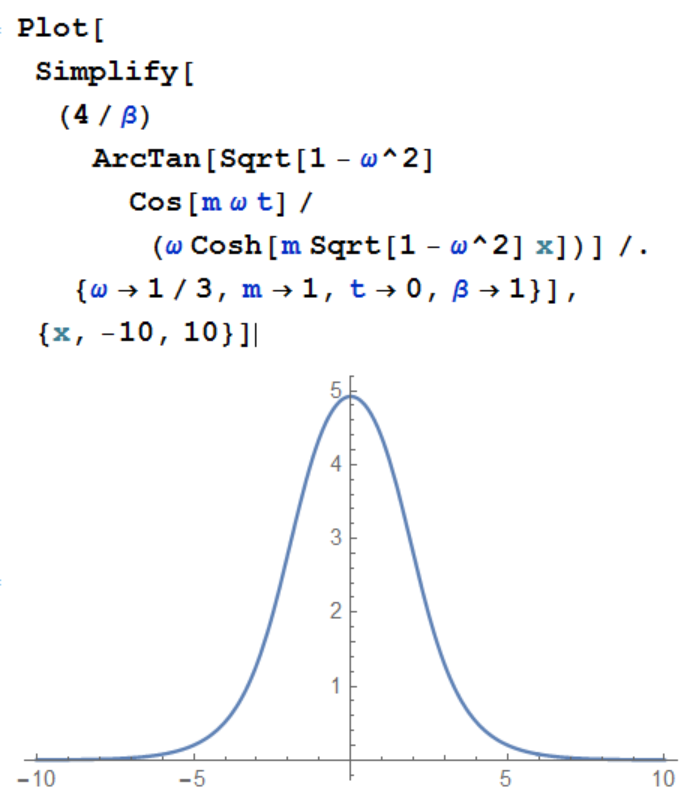}\includegraphics[width=5cm]{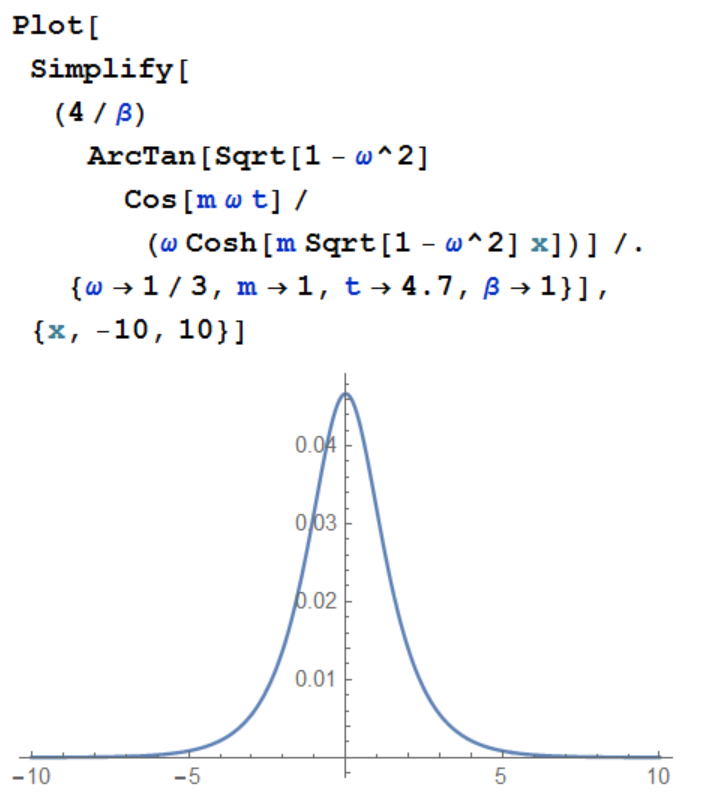}\includegraphics[width=5cm]{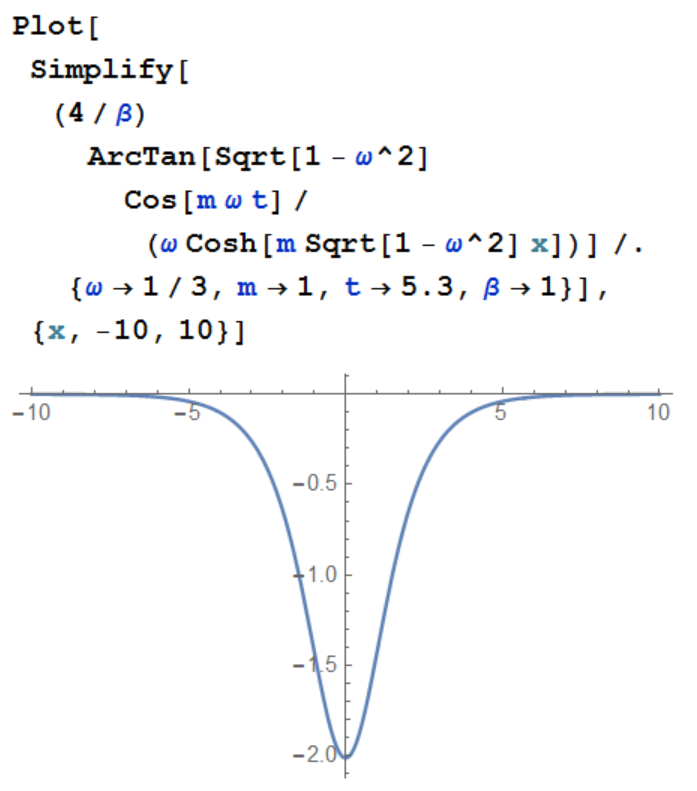}
\caption{The non-translating Sine-Gordon breather in suitable units and for a chosen value of $\omega$, as time evolves (snapshots at three subsequent times $t=0$, $t=4.7$ and $t=5.3$, respectively). Please notice the zoom on the vertical axis.}
\label{sol1}
\end{figure}

The (non-translating) {\it breather} is another solution, given by
\begin{eqnarray}
\phi = \frac{4}{\beta} \arctan \frac{\sqrt{1-\omega^2} \cos (m\omega t)}{\omega \cosh (m\sqrt{1-\omega^2}x)}, \qquad \omega \in [0,1].
\end{eqnarray}
The breather oscillates up and down on the spot (because we have set the velocity of translation to zero) - see fig. \ref{sol1}.
The breather therefore is an ``eigenstate'' of the map $\phi \to - \phi$, in the sense that $-\phi_{breather}$ is a breather oscillating in phase opposition with respect to the original breather (still a solution of course, and, although technically a different solution, for all practical purposes considered as ``the same solution'').

An important remark is that two classical kinks (solitons) repel each other, as can be ascertained from looking at the time evolution of the two-kink solution\footnote{See for instance {\it https://people.maths.ox.ac.uk/trefethen/pdectb/sine-gordon2.pdf} for a discussion on this point.}. The same happens to two antikinks (antisolitons). This will be important in the following, since it will be connected with the fact that the quantised soliton and antisoliton will behave like fermions\footnote{Subtle issues related to this identification will be mentioned in the Conclusions.}. 

{\it Exercise [30 minutes' work using Mathematica]: verify that the soliton, antisoliton and breather profiles have a finite energy $E = \frac{1}{2} {\dot{\phi}}^2 +\frac{1}{2} {\phi'}^2 + \frac{m^2}{\beta^2}(1 - \cos \beta \phi)$, in the case $\beta=m=1$, $v=a=0$, $\omega=\frac{1}{3}$.}

As described in \cite{AleDurham}, integrability allows for a systematic way to find solutions by means of the classical inverse scattering method. One can then obtain multiple travelling profiles which scatter preserving their individual identity. This means that the set of stable classical solutions, in addition to the infinitely many ``vacua'' where the field is constant and equal to one of the minima of the potential, consist of ``superpositions'' (not in the linear sense) of arbitrarily many elementary solutions of the types described above, each travelling along the line and scattering ({\it i.e.}, classically, spending a certain amount of time-delay when passing one another) maintaining their individuality of profiles (akin to multiparticle states with definite momenta on a line). This picture finds a clear correspondence in the factorised scattering theory of the quantised version of these elementary profiles, that is, the quantum particles of the spectrum.

\subsubsection{Lax pair and classical inverse scattering}

The Lax pair for the Sine-Gordon equation is given by \cite{AleDurham}
\begin{eqnarray}
&&L(u) = i\begin{pmatrix}\frac{\beta}{4} \dot{\phi}&\frac{m}{4} u e^{i \frac{\beta}{2} \phi} - \frac{m}{4 u} e^{-i \frac{\beta}{2} \phi} \\\frac{m}{4} u e^{-i \frac{\beta}{2} \phi} - \frac{m}{4 u} e^{i \frac{\beta}{2} \phi}&-\frac{\beta}{4} \dot{\phi} \end{pmatrix},\nonumber\\
&&M(u) = i\begin{pmatrix}\frac{\beta}{4} \phi'&-\frac{m}{4} u e^{i \frac{\beta}{2} \phi} - \frac{m}{4 u} e^{-i \frac{\beta}{2} \phi} \\-\frac{m}{4} u e^{-i \frac{\beta}{2} \phi} - \frac{m}{4 u} e^{i \frac{\beta}{2} \phi}&-\frac{\beta}{4} \phi' \end{pmatrix},\label{LSG}
\end{eqnarray}
where $\dot{\phi} = \partial_t \phi$ and $\phi' = \partial_x \phi$. The spectral parameter $u$ is an arbitrary complex variable, as discussed at the beginning of these lectures.

In \cite{AleDurham} one can find a description of the classical inverse scattering method to obtain solutions of the Sine-Gordon equation, based on the Gelfand-Levitan-Marchenko equation - whose starting point is the Lax pair. One then recovers in particular the soliton and antisoliton profiles described in the previous section by specifying a certain set of {\it classical scattering data} - which are not the quantum scattering matrices, but rather encode the spectral decomposition of a specific differential operator associated with the Lax pair. 

{\it Exercise [20 minutes' work using Mathematica]: verify that (\ref{LSG}) is a Lax pair for the Sine-Gordon equation, and observe first-hand how the spectral parameter decouples from the on-shell data.} 

Lax pairs do in fact belong to Lie algebras \cite{AleDurham} - in the case of Sine-Gordon the $2\times 2$ matrices $L,M$ displayed above are antihermitian. This has a connection with the quantum group symmetry of the quantised version of the model - in this case $U_q\big(\mathfrak{su}(2)\big)$, which we will describe later on. Typically the quantum group reduces to the classical Lie algebra when a suitable parameter (playing the role of $\hbar$) is sent to $0$. 

\subsection{Quantum aspects}

Perturbatively one can establish that the dimensionless coupling $\beta$ is not renormalised, while the dimensional coupling $m^2$ is. The Sine-Gordon Lagrangian is superficially super-renormalisable for $0<\beta^2 < 8 \pi$ and non-renormalisable for $\beta^2 > 8 \pi$. One can see this by looking at the scaling dimension of the cosine term regarded as a perturbation of a free boson (see further down in this section). All the UV divergences in the region $0<\beta^2 < 8 \pi$ can be removed and the dimensional coupling $m^2$ gets multiplicatively renormalised. No wave-function renormalisation for the elementary field is necessary. 

Since $\beta$ is not renormalised, we can fix it to a value of our choice. We will restrict ourselves to $0<\beta^2<8\pi$ for the purposes of this notes. This is called the {\it massive phase} since it has massive soliton and antisoliton excitations. The phase of the theory for $\beta^2 >8 \pi$ is the {\it massless phase} with massless soliton and antisoliton excitations - as we will shortly see, in this regime the cosine perturbation is irrelevant. For comments on the transition between the two regimes see for instance \cite{DD}.

As we discussed in section \ref{exa}, the theory being integrable allows for the application of exact methods which by-pass in principle the perturbative regime and rely only on the structural algebraic properties of the model. Perturbation theory and semiclassical arguments are used as a check of the exact formulas. Moreover, the perturbative expansions also fix some of the parameters which are not constrained when one proceeds purely axiomatically - for instance, the actual renormalised value of the mass can only be ascertained via traditional field-theory methods. 

The renormalisation of the Sine-Gordon model is treated in many classic papers. Reference \cite{FaberIvanov} in particular provides information on the relationship between the dimensional coupling $m$ and the renormalised (running) parameter $m = Z_0 m_r$, with $Z_0$ containing the infinities - which one can tame with a UV cutoff\footnote{It is conceivable that a hard UV cutoff \cite{FaberIvanov} might have the advantage of keeping the theory in $1+1$ dimensions, as opposed for instance to dimensional regularisation. This might retain some of the desired features of integrability. Later on, when we shall introduce the chiral decomposition of the Thirring model, remaining in $1+1$ dimension will have another added bonus. We thank Alexander Ochirov and Dmitrii Riabchenko for discussions on this point. See also \cite{Jarah} for the latest work in a recent important series on the topic.} $\Lambda$. After elimination of the infinities and removing the cutoff, \begin{eqnarray}
m_r^2(M^2) = m_{meas}^2\Big(\frac{M^2}{m^2_{meas}}\Big)^{\frac{\beta^2}{\beta^2 + 8 \pi}}.\label{Fabb}
\end{eqnarray}
$M$ is the {\it scale} and $m_{meas}$ is taken as the {\it reference scale} of the ``measurement'' and also as the physical value of the parameter ``measured'' at the reference scale. 

The relationship between  the renormalised soliton mass $m_{sol}$ (discussed in the next section) and the renormalised dimensional coupling is \cite{ZamoMass,LukZam0,Feverati} (see also \cite{DD})
\begin{eqnarray}
\frac{m_r^2}{2 \beta^2} = m_{sol}^{2 - \frac{\beta^2}{4\pi}} \frac{\Gamma(\frac{\beta^2}{8\pi})}{\pi \, \Gamma(1-\frac{\beta^2}{8\pi})}\Bigg[\frac{\sqrt{\pi} \Gamma(\frac{1}{2}+\frac{\xi}{2\pi})}{2 \Gamma(\frac{\xi}{2\pi})}\Bigg]^{2-\frac{\beta^2}{4\pi}} \to_{\beta \to 0} \, \frac{m_{sol}^2 \beta^2}{128}, \qquad \xi = \frac{\beta^2}{8} \frac{1}{1 - \frac{\beta^2}{8\pi}}.\label{ZLu}
\end{eqnarray}

%We can check that, as $\beta \to 0$, we have
%\begin{eqnarray}
%\frac{m_r^2}{\beta^2} (\cos \beta \phi - 1) \to - m_{sol}^2 \frac{\beta^4 \phi^2}{64}.
%\end{eqnarray}
%We will see later on that the spectrum contains breathers, the first of which is the quantum of the field $\phi$ itself. The mass $M_1$ of the first breather, which can be evinced from the general formula (see (\ref{spec}) later on), has a small-$\beta$ expansion whose leading term is $M_1 \to \frac{m_{sol} \beta^2}{8}$, therefore at small $\beta$ we deduce
%\begin{eqnarray}
%\frac{m_r^2}{\beta^2} (\cos \beta \phi - 1) \to - M_1^2 \phi^2.
%\end{eqnarray} 
%This identifies the classical limit of the field-quantum mass $M_1$ with the classical parameter $\sqrt{2} m$.

Perturbative and semiclassical evidence for the factorisation of the $S$-matrix in Sine-Gordon (and in the Thirring model as well) is found in \cite{Are}. We remind that semi-classical phase-shifts are related to time delays in soliton scattering \cite{Woo}.

{\it Further reading [two day's study]: become familiar with section 4 in \cite{Factorised}.} 

In the massive phase, the Sine-Gordon theory can be understood as a perturbation of the free-boson CFT by the operator $\cos \beta \phi$ (more references will be given in the Conclusions). The perturbing (composite) operator has a quantum-mechanical scaling dimension $[length]^{-\frac{\beta^2}{4 \pi}}$, therefore the parameter $m^2$ acquires a quantum-mechanical mass-dimension of $2 - \frac{\beta^2}{4\pi}$, as can clearly be seen by (\ref{ZLu}). This means that the perturbation is relevant for $0<\beta^2<8 \pi$. 

This triggers a renormalisation-group flow from the massless UV fixed point (the free boson indeed, with central charge $c=1$) to the trivial IR fixed point with $c=0$ with no excitations.  Such a flow can be described by the thermodynamic Bethe ansatz (TBA), which provides a (particular) $c$-function $c(m_{sol} R) = \frac{6 R E_0(R)}{\pi}$ (``running central charge''), where $E_0(R)$ is the ground-state energy of the theory on a circle of radius $R$ as computed by the TBA, and $m_{sol}$ is the soliton mass (see next section). The parameter $R$ works as a scale along the flow \cite{FenSal}, such that (in unitary theories) $c(0)=c_{UV}$ and $c(\infty) = c_{IR}$ \cite{FenSal2}, and $c_{UV}> c_{IR}$. For a study away from the massive phase the reader can consult for instance \cite{Defenu}.  

The Sine-Gordon model at special values of $\beta$ reduces to other known integrable models or presents particular extra features:
\begin{enumerate}
\item We will have ample way of motivating that at the value $\beta^2 = 4 \pi$ (self-dual radius of the compact boson $r_{bos} = \frac{4\pi}{\beta^2}=1$, see for instance \cite{TongNotes}) the model develops a free-fermion behaviour. 
\item At $\beta^2 = \frac{16 \pi}{3}$ (which is in the so-called {\it repulsive} regime, see the next section) the model develops ${\cal{N}}=2$ supersymmetry \cite{iFendley}. In the massless limit the theory at this value of $\beta$ reduces to a free boson compactified on a special radius $r_{bos} = \frac{4\pi}{\beta^2} = \frac{3}{4}$(see for instance \cite{iFendley}).
\item At $\beta^2 = 8 \pi$ (the perturbation has become marginal) the model maps to the $\mathfrak{su}(2)$ Gross-Neveu model, which is known to exhibit Yangian symmetry \cite{Loebbert}.
\end{enumerate}
The reader is also invited to consult \cite{Freedman:1988xh}.

\subsubsection{Soliton $S$-matrix and dressing phase}

The main reference for this section is \cite{Diego}. 

The spectrum of the quantised Sine-Gordon theory is built out of a soliton and an antisoliton, which may or may not form a bound state. In the semiclassical (small $\beta^2$) approximation the mass of the soliton and antisoliton is \cite{DashenHasslacherNeveu,FaberIvanov} 
\begin{eqnarray}
m_{sol} = \frac{m_{meas}}{\xi} = \frac{8 m_{meas}}{\beta^2} - \frac{m_{meas}}{\pi}.\label{Da}
\end{eqnarray}
The mass (rest-frame energy) was originally calculated in  \cite{DashenHasslacherNeveu}. Formula (\ref{Da}) consists of the classical contribution plus the first quantum correction from small oscillations, with counterterms suitably subtracted. Both terms have been evaluated at the reference scale $m_{meas}$ \cite{FaberIvanov}. Considering that at leading order at small $\beta^2$ we have from (\ref{Fabb}) $m_r^2 \sim m_{meas}^2$, (\ref{Da}) gives $m_{sol} \to \frac{8 m_{meas}}{\beta^2} \sim \frac{8 m_{r}}{\beta^2}$ at small $\beta^2$, which fits with (\ref{ZLu}).

The bound state is present in the spectrum when we are in the so-called {\it attractive regime} $0<\beta^2 < 4\pi$, while it is absent in the {\it repulsive} regime $\beta^2 > 4 \pi$. The soliton and antisoliton are the ``fundamental'' quantum excitations of the model, and correspond to the quantisation of the classical soliton and antisoliton profiles. They are {\it not} the excitations (quanta) of the field $\phi$ itself. In fact, the quantum of the field $\phi$ has rather more to do with the soliton-antisoliton bound state - not being literally the same but being related, in a way which might become slightly clearer when we shall introduce the duality with the Thirring model, see (\ref{refin}) and (\ref{that}). 

The soliton-antisoliton $S$-matrix is explicitly constructed in \cite{ZamoSG,Factorised}, see \cite{Diego} for a review. Here we simply report its salient features. In the ordered basis $(s,\bar{s})$, where $s$ stands for the quantum soliton state and $\bar{s}$ for the antisoliton, the $S$ matrix takes the form
\begin{eqnarray}
S_{SG} = \begin{pmatrix}S&0&0&0\\0&S_T&S_R&0\\0&S_R&S_T&0\\0&0&0&S\end{pmatrix},\label{upper}
\end{eqnarray}
where
\begin{eqnarray}
S = S(\theta), \qquad S_T = \frac{\sinh \frac{\pi \theta}{\xi}}{\sinh \frac{\pi (i\pi-\theta)}{\xi}} S(\theta), \qquad S_R = \frac{i \sin \frac{\pi^2}{\xi}}{\sinh \frac{\pi (i\pi-\theta)}{\xi}} S(\theta).\label{lower}
\end{eqnarray}
The labels $T$ and $R$ stand for {\it transmission} and {\it reflection}, respectively, and correspond to the soliton and antisoliton ``flavours'' of the scattering particles being either preserved or exchanged. The matrix $S_{SG}$ is written in the basis $\{|s\rangle \otimes |s\rangle,|s\rangle \otimes |\bar{s}\rangle,|\bar{s}\rangle \otimes |s\rangle,|\bar{s}\rangle \otimes |\bar{s}\rangle$\} - it is therefore written as a quantum-group $R$-matrix (see section \ref{qgroup}). This means that for instance the $S_T$ entries send $|s(\theta_1)\rangle \otimes |\bar{s}(\theta_2)\rangle \to |s(\theta_1)\rangle \otimes |\bar{s}(\theta_2)\rangle$ and $|\bar{s}(\theta_1)\rangle \otimes |s(\theta_2)\rangle \to |\bar{s}(\theta_1)\rangle \otimes |s(\theta_2)\rangle$. The physicists' way of writing the $S$-matrix would instead entail that $S_T$ is the amplitude for the processes $|s(\theta_1)\rangle \otimes |\bar{s}(\theta_2)\rangle \to |\bar{s}(\theta_2)\rangle \otimes |s(\theta_1)\rangle$ and $|\bar{s}(\theta_1)\rangle \otimes |s(\theta_2)\rangle \to |s(\theta_2)\rangle \otimes |\bar{s}(\theta_1)\rangle$ (making the image of ``transmission'' more manifest). The conservation of the total soliton number (topological charge) sets to zero any other process (entries of the $S$-matrix). In fact the upper $1 \times 1$ block of (\ref{upper}) is the charge $+2$ super-selection sector, the central $2 \times 2$ block the charge $0$ sector, and the lower $1 \times 1$ block the charge $-2$ sector. Such sectors would be completely independent, if it were not for the crossing-symmetry requirement which mixes them. In addition, the manifest democracy between the soliton and the antisoliton provides an intuitive justification as to why some of the non-zero entries turn out to be equal, while (\ref{lower}) comes from imposing some of the scattering ``axioms''.

Notice that the parameter $\xi$ is ill-defined for $\beta^2 = 8 \pi$. This is a delicate point and is discussed in \cite{Coleman}. Coleman argued that the energy density of the theory becomes unbounded below if $\beta^2$ exceeds $8 \pi$. He argued this by means of a perturbative analysis starting from the trivial vacuum $\phi=0$. It was later remarked \cite{Factorised,FaberIvanov} that the theory can be defined for values of $\beta^2$ larger than $8 \pi$, provided one perturbatively expands around a non-trivial vacuum \cite{FaberIvanov}. We shall always assume $\beta^2 < 8 \pi$ in these notes.

We refer to \cite{Diego} for the entire step-by-step procedure leading to (\ref{lower}), which is not yet the complete story. At the end of such a procedure, the overall factor $S(\theta)$ ends up having to solve two conditions dictated by braiding unitarity and crossing symmetry. The two conditions read 
\begin{eqnarray}
\label{unicross}
S(\theta)S(-\theta) = 1, \qquad S(\theta) = \frac{\sinh \frac{\pi (i\pi - \theta)}{\xi}}{\sinh \frac{\pi \theta}{\xi}} S(i\pi-\theta)
\end{eqnarray}
and they are actually sufficient to determine a {\it minimal} solution:
\begin{eqnarray}
\label{Zamofactor}
S(\theta) = - \prod_{k=0}^\infty \frac{\Gamma\big(1+(2k+1)\frac{\pi}{\xi} - i \frac{\theta}{\xi}\big)\Gamma\big(1+2k\frac{\pi}{\xi} + i \frac{\theta}{\xi}\big)\Gamma\big((2k+1)\frac{\pi}{\xi} - i \frac{\theta}{\xi}\big)\Gamma\big((2k+2)\frac{\pi}{\xi} + i \frac{\theta}{\xi}\big)}{\Gamma\big(1+(2k+1)\frac{\pi}{\xi} + i \frac{\theta}{\xi}\big)\Gamma\big(1+2k\frac{\pi}{\xi} - i \frac{\theta}{\xi}\big)\Gamma\big((2k+1)\frac{\pi}{\xi} + i \frac{\theta}{\xi}\big)\Gamma\big((2k+2)\frac{\pi}{\xi} - i \frac{\theta}{\xi}\big)}.
\end{eqnarray}
The attribute of minimal always refers to having the smallest number of poles and zeroes in the physical strip $\mbox{Im} \, \theta \in (0,\pi)$. 

{\it Exercise [2 days' work]: Prove that (\ref{Zamofactor}) satisfies the second relation (\ref{unicross}) by using known identities of the Gamma function.}

{\it Guided exercise [2 hours' work]: Reproduce the two different demonstration given in \cite{Diego} - pages 33-35.}

There exists a well-known integral representation for $S(\theta)$:
\begin{eqnarray}
S(\theta) = -\exp \Bigg[-i\int_0^\infty \frac{dt}{t}\frac{\sinh \Big(t(\pi - \xi)\Big)}{\sinh (\xi t) \cosh (\pi t)} \sin (2 \theta t)\Bigg], \label{Malm}
\end{eqnarray}
see also \cite{Karoetal}. More precisely we can say that (\ref{Zamofactor}) is the analytic continuation of (\ref{Malm}) to the whole complex $\theta$-plane.

{\it Exercise [1 day's work]: obtain (\ref{Malm}) from (\ref{Zamofactor}) by using the so-called {\it Malmst\'en} integral representation of the Gamma function:
\begin{eqnarray}
\Gamma(x) = \exp \int_0^\infty \frac{dt}{t}\Big[(x-1)e^{-t} + \frac{e^{-tx}-e^{-t}}{1-e^{-t}}\Big],\qquad \mbox{\rm Re}(x) >0.
\end{eqnarray}
Identify the region in the $\theta$-plane where (\ref{Malm}) is valid.
}

Being a phase factor for real values of $\theta$, as required by physical unitarity, and being an overall factor of the $S$-matrix, a function such as $S(\theta)$ is often referred to (particularly in the AdS/CFT integrability community) as a {\it dressing factor} and its $\log$ as a {\it dressing phase}. 

Notice that the overall minus sign in $S(\theta)$ is at this stage completely arbitrary, since the two conditions (\ref{unicross}) are quadratic in the dressing factor. It is however {\it very important} that this minus sign is put there by hand, for reasons which will become clear in section \ref{Equivalence}. In fact, the condition at equal rapidities (equal momenta) of $S(0)=-1$ when the two identical scattering particles' bare statistics is bosonic, effectively turns them into behaving like fermions, which is what will happen here. As pointed out by Zamolodchikov, braiding unitarity for identical particles implies that their scattering amplitude satisfies $S(0)^2=1$ hence $S(0)=\pm 1$. This argument applies when the two particles scatter diagonally, {\it i.e.} with no transformation of their internal degrees of freedom - otherwise braiding unitarity would involve a sum of terms corresponding to the different channels. When insterted in the Bethe wave-function, this either retains ($S(0)=+1$) or changes ($S(0)=-1$) the bare statistics. An easy way of seeing this is by resorting to a different and simpler model, the non-linear Schroedinger (or Lieb-Liniger) model. In that case, the two-particle Bethe wave-function can be written as
\begin{eqnarray}
&&|p_1,p_2\rangle \equiv \int_{-\infty}^\infty\int_{-\infty}^\infty dx_1 dx_2 \Big[\Theta(x_1 - x_2) + S_{LL}(p_1,p_2) \Theta(x_2 - x_1) \Big] e^{i p_1 x_1 + i p_2 x_2} \, a^\dagger(x_1) a^\dagger(x_2)|0\rangle \nonumber \\\label{org}
\end{eqnarray}
where $p$ is the particle momentum, $a^\dagger$ are the creation operators associated with the perturbative excitations of the field (in that case a complex boson with Galileian invariance), and $|0\rangle$ is actually the perturbative vacuum annihilated by all the $a$'s. The state (\ref{org}) is an exact stationary state - solution of the exact time-independent Schroedinger's equation in the sector where the particle number $N$, which is conserved, is $N=2$. It is an {\it in} state for $p_1<p_2$ and an {\it out} state for $p_1>p_2$, with the {\it in} and {\it out} states being rightfully related by the $S$-matrix $S_{LL}$  - see \cite{Evgeny2}, around formula (2.10). The state is in principle delta-function normalisable (although we have not bothered reporting the correct normalisation here). The eigenvalue is $E = p_1^2 + p_2^2$ (Galileian). The Lieb-Liniger model is simpler than Sine-Gordon, in that one can work with the perturbative modes (the Fourier modes of the field) and with relatively little effort construct exact eigenstates of the quantum Hamiltonian using those modes, by organising them in the fashion of (\ref{org}) and its $N>2$ generalisations. The $S$-matrix is Galieian-invariant: $S_{LL}(p_1,p_2) = S_{LL}(p_1-p_2) =\frac{p_1 - p_2 + i \zeta}{p_1 - p_2 - i \zeta}$ for some coupling $\zeta$. Clearly $S_{LL}(0)=-1$. The wave-function (\ref{org}) vanishes for $p_1=p_2$ precisely because the creation operators $a^\dagger$ are bosonic and $S_{LL}(0)=-1$: we have an effective exclusion principle. In the strong $\zeta$ regime the scattering bosons have an $S$-matrix that tends to $-1$ for any value of the momenta: the model has the characteristic behaviour of free fermions (impenetrable bosons). What we have described is a completely non-perturbative phenomenon: trivially a value of the $S$-matrix of $-1$ cannot be expanded as $1 + small$ (perturbation around free particles), as $-1$ is always a finite distance away from $1$. If $\zeta$ is finite, then in perturbation theory the particles are bosons, since $S(p_1-p_2) \sim 1 + 2 \sum_{n=1}^{n_0} \big(\frac{i \zeta}{p_1-p_2}\big)^n$ for some large $n_0$, hence $S(0)=1$ at any order in perturbation theory. However the exact resummation reveals their fermionic behaviour, since the sum satisfies, as we have seen, $S(0)=-1$. It is interesting to consider that, where we to write (\ref{org}) as
\begin{eqnarray}
|p_1,p_2\rangle = {\cal{A}}^\dagger_{p_1} \, {\cal{A}}^\dagger_{p_2} |0\rangle, 
\end{eqnarray} 
we would have to conclude that the operator ${\cal{A}}^\dagger_{p}$ satisfies
\begin{eqnarray}
\big({\cal{A}}^\dagger_{p}\big)^2|0\rangle=0, 
\end{eqnarray}
which of course has the feeling of creating fermionic-type objects. Such operators can be constructed and lead one naturally into the larger framework of Faddeev-Zamolodchikov operators and the algebraic Bethe ansatz (see later sections). We refer to \cite{Evgeny,Evgeny2} for the full detail - see for instance formulas (3.81) and (3.83) in \cite{Evgeny2}, taken at $k=k'$.

Let us conclude this section by remarking that the infinite product of Gamma functions displayed in the Sine-Gordon dressing factor converges thanks to the following theorem\footnote{We also thank Sergey Frolov for discussions on this point.} - \cite{Kulkarni} pag. 178. 

{\it Theorem} Given $N_1$ and $N_2$ complex numbers $\alpha_1,...,\alpha_{N_1}$ and $\beta_1,...,\beta_{N_2}$, respectively, the infinite product
\begin{eqnarray}
\prod_{n=1}^\infty \frac{\Gamma(n-\alpha_1)...\Gamma(n-\alpha_{N_1})}{\Gamma(n-\beta_1)...\Gamma(n-\beta_{N_2})}
\end{eqnarray}
converges if and only if
\begin{enumerate}
\item $N_1=N_2=N$ and
\item $\sum_{i=1}^N\alpha_i = \sum_{i=1}^N \beta_i$ and
\item $\sum_{i=1}^N\alpha_i^2 = \sum_{i=1}^N \beta_i^2$.
\end{enumerate}

Representing the dressing factor as a convergent infinite product of Gamma functions is ideal to identify poles and zeros, simply because the Gamma function is never zero and it has poles at negative-integer and zero argument. 

{\it Exercise [one afternoon's work]: list all the poles and zeroes of $S(\theta)$, $S_T(\theta)$ and $S_R(\theta)$ in the whole complex plane, with their respective order.}

Notice that remarkably\footnote{We thanks Chawakorn Maneerat for pointing this out to us.} the Sine-Gordon $S$-matrices (see also the next section) do not carry a dependence on the renormalisation scale, which is confined in the mass parameter and hence in the particles' dispersion relation.

\subsection{Breather $S$-matrix, mixed $S$-matrix}

The poles in the fundamental $S$-matrix which can give rise to bound states are only those within the physical strip Im $\theta \in (0,\pi)$. In the attractive regime $\xi<\pi$ the bound state spectrum was found by \cite{DashenHasslacherNeveu} via quantisation of the classical two-soliton solution:
\begin{eqnarray}
M_k = 2 m_{sol} \sin \frac{k \xi}{2}, \qquad k = 1,2,...<\frac{8\pi}{\beta^2}-1 = \frac{\pi}{\xi}.
\label{spec}
\end{eqnarray}
Formula (\ref{spec}) is believed to be exact (see also comments in \cite{Factorised}). 
As $\beta \to 0$ the breather spectrum becomes continuum, which is the classical result. The small-$\beta$ expansion of the inverse expression $m_{sol} = \frac{M_1}{2 \sin \frac{\xi}{2}}$ reads
\begin{eqnarray}
m_{sol} \sim \frac{8M_1}{\beta^2}-\frac{M_1}{\pi},
\end{eqnarray}
which matches the semiclassical result (\ref{Da}) if we identify \underline{at small coupling} the first breather with the field quantum, in the sense that $M_1 \to m_{meas}\sim m_r$ as $\beta \to 0$.

{\it Guided exercise [4 hours's work]: reproduce the discussion in \cite{Patrick} - pages 19-21.}

{\it Further reading [4 hours]: explore the Coleman-Thun mechanism \cite{Patrick} - pages 21-23 and \cite{ColemanThun}.}

As an example, the exact $S$-matrix of two breathers of mass $M_1 = 2m_{sol}\sin \frac{\xi}{2}$ is given by \cite{KarowskiWeisz}
\begin{eqnarray}
S_{11} = \frac{\sinh \theta + i \sin\xi}{\sinh \theta - i \sin\xi}. \label{brrr}
\end{eqnarray}
This amplitude has simple poles at $\theta = i\xi, i (\pi -\xi)$. They correspond to the direct ($s$-channel) and the crossed ($t$-channel) appearance of a $k=2$ breather: for instance, computing the Mandelstam invariant in the direct channel returns
\begin{eqnarray}
(p_1+p_2)^2 = 2 M_1^2 (1+\cosh \theta_{pole})=2 M_1^2 \Big(1+\cosh i\xi \Big) = 4 M_1^2 \cos^2 \frac{\xi}{2} = 4 m_{sol}^2 \sin^2 \xi = M_2^2,
\end{eqnarray} 
where we have used (\ref{spec}) twice. These poles fall within the physical strip Im $\theta \in (0,\pi)$. It is noteworthy that the amplitude (\ref{brrr}) is a CDD factor\footnote{It is also of identical functional form to the $S$-matrix of the single bosonic excitation of the so-called {\it Sinh-Gordon} model, see for instance \cite{AhnDelfinoMussardo} - we thank one of the LonTI attendees for pointing this out to us.}, see (\ref{cDd}).

{\it Exercise [absorbing as much time as you wish, using hints from \cite{Patrick,Diego}]: have fun doing the same invariant-mass analysis, as we have just done, for the poles of the various soliton amplitudes.}

Notice that by being a bound state of a soliton and an antisoliton the breather has zero topological charge and it is a boson, as befits the connection with the fundamental quantum of the scalar field $\phi$. We also remark that at the onset of the repulsive regime all bound states cease to exist, meaning that they are all unstable against the decay into a soliton and an antisoliton. As the coupling increases, and as it is well visible from (\ref{spec}), the breathers actually disappear one by one from the spectrum (each in turn becoming unstable against decay) every time $\frac{8\pi}{\beta^2}-1$ passes an integer, until finally only the soliton and the antisoliton are left \cite{DashenHasslacherNeveu}.  

There is a certain degree of ambiguity in determining which of the particles is in fact the ``fundamental'' quantum - an aspect which has been deemed {\it nuclear democracy} \cite{Chew}. This corresponds to the idea that bound states can be conceptually perceived as being as ``fundamental'' as their constituents, and that if one uses a different theory they might be granted a field of their own. The paper \cite{DashenHasslacherNeveu} has a great deal of discussion around the relation between the lowest-lying breather and the quantum of $\phi$. We will see later on that this principle of nuclear democracy applies quite universally. (Anti)solitons are collective excitations of $\phi$, but in a dual picture they will be the fundamental fermionic-like objects and will have bosonic bound states - on the other hand, the theory describing these bosonic bound states is the theory of the field $\phi$. In the end there simply exist different field-theory descriptions of the physics, with different fields as the basic variables, and asking which field is ``fundamental'' and which is ``composite'' ends up being rather besides the point. 

A more complete account of the $S$-matrices of Sine-Gordon bound states can be found in \cite{KarowskiThun} - see also \cite{ZamoITEP}.

There is a conserved ``parity'' charge for breathers corresponding to the $\phi \to - \phi$ symmetry. The breathers can be seen to be eigenstates of this flip operation, and have charge parity $(-)^k$.  

\subsection{Sine-Gordon and the XXZ spin-chain}

There is a very close connection between the Sine-Gordon model and the so-called XXZ spin-chain - a famous model of magnetism in lower dimensions. This is the discrete model of spins in the $\frac{1}{2}$ (fundamental) representations of $\mathfrak{su}(2)$, each pinned down at one site of a one-dimensional lattice (chain) so as to be each distinguishable from any other, with quantum-mechanical Hamiltonian
\begin{eqnarray}
H_{XXZ} = J \sum_{i=1}^N \big[\sigma_{1,i} \sigma_{1,i+1} +\sigma_{2,i} \sigma_{2,i+1} + \Delta \, \sigma_{3,i} \sigma_{3,i+1}\big], 
\end{eqnarray}
where $N$ is the number of lattice sites \cite{Ana2}. We impose periodic boundary conditions - the site $N+1$ is the site $1$. The parameter $J$ is the {\it exchange coupling}. The name XXZ derives from the fact that the $\sigma_3 - \sigma_3$ coupling is in general different from the other two (the {\it anisotropy parameter} $\Delta$ not being necessarily equal to $1$, which would reduce the theory to the Heisenberg model). The chain has nearest-neighbour interactions only - the Hamiltonian density involves only adjacent sites ($i$ and $i+1$).
 
This quantum-mechanical spin-chain model is integrable - it directly admits a quantum $L$-operator from which a quantum monodromy matrix can be built, and 
$N$ commuting conserved local charges can be obtained from it (in parallel with what happens for classical field theories). The fact that a quantum field theory and a spin-chain model are so intimately connected is largely due to the fact that they share a common underlying integrable structure (encoded in a special quantum group which we will review in the next section). It is however still a rather remarkable realisation. To do justice to such relationship would require a longer treatment which goes beyond the scope of these lectures, therefore we will just basically provide a small guide the literature, while we will focus a bit more on the quantum-group basics.  

A link between the two models is described in \cite{FaddeevHow} - eq.s (420)-(449) there. One can construct a rather involved way of obtaining the Sine-Gordon field-equation from a formal continuum limit of the RTT relations (see later) associated with the XXZ chain, when the $L$-operator associated with the latter is re-expressed in terms of suitable non-commuting variables. The quantum $L$-operator can be considered as a quantum analogue of the classical Lax matrix $L$. The non-commuting variables one obtains satisfy exchange relations which are reminiscent of those defining the so-called {\it quantum plane} \cite{Kassel}, already anticipating the underlying connection with $U_q\big(\mathfrak{su}(2)\big)$ - see the next subsection. The XXZ chain therefore can be seen as a lattice regularisation of quantum Sine-Gordon.

In \cite{FaddeevTirkkonen} - section 3.1 there - a link is then drawn between the L-operator of a lattice regularisation of Sine-Gordon and the XXZ L-operator. The two L-operators are essentially seen as arising from different representations of the fundamental quantum group (RTT) relations, in particular they share the same (trigonometric) $R$-matrix (\ref{erreb}).

A connection has also been established by Destri and De Vega \cite{DDV} - see also \cite{Feverati} - using the general framework of the six-vertex model.

Finally, we recommend chapter XVIII of \cite{Miranda} for a discussion of the relationship between the XXZ spin-chain and the Sine-Gordon model in the context of bosonisation, and the recent \cite{Murugan:2018fdj}.

\subsection{The quantum group $U_q\big(\mathfrak{su}(2)\big)$\label{qgroup}}

Integrable systems are tightly connected with the mathematical theory of quantum groups, which provides the unified framework to describe them. From the quantum-group viewpoint, there is no difference between a quantum field theory and a quantum-mechanical spin-chain, as long as their quantum conserved charges define the same {\it Hopf algebra} (which for our purposes is a synonym of {\it quantum group}). 

For a review of Hopf algebras we refer for example to \cite{Kassel}. Here we simply recall the definition of the Hopf algebra $U_q\big(\mathfrak{su}(2)\big)$. This Hopf algebra is the deformation of the universal enveloping algebra $U\big(\mathfrak{sl}(2)\big)$ characterised by a (in principle complex) deformation parameter $q$. It can be presented as the algebra generated by arbitrary polynomials in the three generators $E,F,H$ subject to the relations ($q \neq \pm 1$)  
\begin{eqnarray}
[H,E]=E, \qquad [H,F]=-F, \qquad [E,F] = \frac{q^{2H} - q^{-2H}}{q-q^{-1}}.
\label{rela}
\end{eqnarray}
Here we adopt the conventions of \cite{Crampe}.
As $q \to 1$, these relations tend to those defining the Lie algebra $\mathfrak{sl}(2)$. One then chooses a suitable real form to obtain $\mathfrak{su}(2)$. The {\it algebra} structure is given by the multiplication of two generators and the unit w.r.t. such multiplication. In the universal envelop we indeed are allowed to write $[a,b] = a.b-b.a$, with $.$ being the multiplication - multiplying generators is in fact the only way we can be allowed to write something like $q^H$, which is formally a power series. The generators act in various representations, which defines the particles of the field theory (or the sites of the spin-chain). The algebra is still a vector space.

In order to qualify as a Hopf algebra, we have to exhibit a coproduct for our algebra. This is a map
\begin{eqnarray}
\Delta: U_q\big(\mathfrak{su}(2)\big) \to U_q\big(\mathfrak{su}(2)\big) \otimes U_q\big(\mathfrak{su}(2)\big).
\end{eqnarray}
The coproduct physically gives the action of the symmetry generators on two particles of a field theory (or two sites of a spin-chain):
\begin{eqnarray}
\Delta(E) = q^H \otimes E +  E \otimes q^{-H}, \qquad \Delta(F) = F \otimes q^{-H} +  q^H \otimes F, \qquad \Delta(H) = H \otimes \mathbbmss{1} +  H \otimes \mathbbmss{1}.
\label{cop}
\end{eqnarray}
The identity $\mathbbmss{1}$ is in a sense the zero-th power of the generators. When $q \to 1$, the coproduct reduces to the standard Leibniz rule for two-particle symmetries. The presence of $q$ consistently {\it deforms} the standard structures.

{\it Exercise [15 minute's work]: verify that (\ref{cop}) is an algebra homomorphism - meaning that replacing every generator by its coproduct preserves all the relations (\ref{rela}). The exponentials of tensor products can be more easily dealt with by using a formal Taylor expansion, and noticing that (for bosonic generators) $a\otimes b \cdot c \otimes d = ac \otimes bd$.}

The coproduct being an algebra homomorphism implies that, given a representation of $U_q\big(\mathfrak{su}(2)\big)$ one can generate more representations by repeatedly applying the coproduct: this is nothing else but tensoring representations - it is the same as the usual composition of spins in quantum mechanics, except for the fact that the quantum mechanical textbook case is always implicitly performed with the standard (Leibniz) coproduct, namely undeformed by $q$. Whenever we wrote
\begin{eqnarray}
\mbox{total spin} = \mbox{spin}_1 + \mbox{spin}_2
\end{eqnarray}
we have always secretly meant (perhaps without realising it)
\begin{eqnarray}
\Delta(\sigma_i) = \sigma_i \otimes \mathbbmss{1} + \mathbbmss{1} \otimes \sigma_i,
\end{eqnarray}
for the $i$-th component of the total spin of two particles.

There is also a counit (mapping the Hopf algebra into the field of coefficients, typically $\mathbbmss{C}$)
\begin{eqnarray}
\epsilon(E) = \epsilon(F) = \epsilon(H) = 0, \qquad \epsilon(\mathbbmss{1})=1,
\end{eqnarray}
which completes the so-called {\it co-algebra} structure. Finally, the antipode is obtained from the knowledge of the other four maps (multiplication, unit, coproduct and counit) \cite{Kassel}, and reads
\begin{eqnarray}
\Sigma(E) = - E, \qquad \Sigma(F) = - F, \qquad \Sigma(H) = - H, \qquad \Sigma(\mathbbmss{1}) = \mathbbmss{1}.
\end{eqnarray} 
The minus sign is reminiscent of what an inverse does to a group element ({\it i.e.} changing sign to a Lie algebra element). Physically, there is a route from the antipode to the notion of antiparticles \cite{BernardLeClair}.  

The $U_q\big(\mathfrak{su}(2)\big)$ Hopf algebra is {\it quasi-cocommutative}, meaning that there exists an $R$-matrix. This is an invertible object satisfying 
\begin{eqnarray}
\Delta^{op}(a) R = R \Delta(a), \qquad a=E,F,H,
\label{quasico}
\end{eqnarray}
where the {\it opposite} coproduct $\Delta^{op}$ is obtained from $\Delta$ by permuting the two factors of the tensor product (for instance $\Delta^{op}(E) = E \otimes q^H + q^{-H} \otimes E$). The relation (\ref{quasico}) says that the coproduct map is not cocommutative (that would be $\Delta^{op} = \Delta$), but it ``almost'' is (the Italian {\it quasi}), since $\Delta^{op}(a) = R \Delta(a) R^{-1}$ for each of the generators $a=E,F,H$. Physically, we should think of $R$ as a relative of a two-particle $S$-matrix, precisely $R = perm \circ S$ with $perm$ the permutation of the two particles. The permutation will be graded, so permuting two fermionic-like objects generates a minus sign. A cocommutative coproduct would reduce (\ref{quasico}) to $[R,\Delta(a)]=0$, which is the ordinary meaning of a local symmetry of the $S$-matrix. Formula (\ref{quasico}) generalises this notion to coproducts which are non-local, i.e. non cocommutative.  

Another great power of Hopf algebras is revealed by the fact that we can solve (\ref{quasico}) in {\it any} representation: the standard solution is - see for instance \cite{Crampe} - 
\begin{eqnarray}
R =  \sum_{n \geq 0} \frac{(q-q^{-1})^n}{[n]_q!}q^{\frac{n(n-1)}{2}}\Big(F q^H \otimes q^{-H} E\Big)^n q^{2\big(H \otimes H\big)},
\label{uni}
\end{eqnarray}
where
\begin{eqnarray}
[n]_q! = [n]_q[n-1]_q... [2]_q[1]_q, \qquad [n]_q =  \frac{{q^n} - q^{-n}}{q - q^{-1}}.\label{uni2}
\end{eqnarray}
This formula is called the {\it universal} $R$-matrix precisely because it does not care about the representation, it works purely based on the commutation relations. It also satisfies the Yang-Baxter equation purely algebraically. 

{\it Exercise [to do once in a lifetime]: verify (\ref{quasico}) using \underline{only} the formula for the universal $R$-matrix (\ref{uni})-(\ref{uni2}) and the defining relations of $U_q\big(\mathfrak{su}(2)\big)$.} 

Once a relation has been established for the three generators, it will follow for any polynomial thereof simply by homomorphism. The striking feature is that now we can use (\ref{uni}) in any of the representations of $U_q\big(\mathfrak{su}(2)\big)$ - each time obtaining a different matrix solution of the Yang-Baxter equation. Each of these different matrices defines a different integrable scattering problem, but all of these problems are still connected with the same underlying algebraic structure. They can also define different integrable models altogether, and those integrable models will still be connected by their algebraic binding. For example, by plugging in the fundamental representation 
\begin{eqnarray}
E = \sigma_+ = \frac{1}{2}(\sigma_1+i\sigma_2), \qquad F = \sigma_- = \frac{1}{2}(\sigma_1-i\sigma_2), \qquad H = \frac{1}{2}\sigma_3,
\label{repa}
\end{eqnarray}
into (\ref{uni})-(\ref{uni2}) one obtains the $R$-matrix 
\begin{eqnarray}
R = \begin{pmatrix}\sqrt{q}&0&0&0\\0&\frac{1}{\sqrt{q}}&0&0\\0&\frac{q-\frac{1}{q}}{\sqrt{q}}&\frac{1}{\sqrt{q}}&0\\0&0&0&\sqrt{q}\end{pmatrix}.
\label{erre}
\end{eqnarray}
Notice that the representation (\ref{repa}) does not depend on $q$ and it is therefore also a representation of $\mathfrak{sl}(2)$. This is accidental, and it is a peculiarity of very special representations such as the fundamental. 

{\it Exercise [2 hours' work]: exploiting the fact that $E$ and $F$ in the fundamental representation are nilpotent matrices, obtain (\ref{erre}) from (\ref{uni}). Then check that (\ref{quasico}) is satisfied in the fundamental representation.}

Most of the ``axioms'' of the $S$-matrix programme can be incorporated into the framework of Hopf algebras, and in this way they become amenable to solutions by methods of representation theory. In fact the theory of quantum groups was originally developed by Drinfeld as a way of \underline{generating} solutions of the Yang-Baxter equation via an algebraic machinery. Such solutions are ready to then become the $S$-matrices of integrable systems, as they naturally satisfy the axioms - in fact all such $S$-matrices descend from the universal $R$-matrix which satisfies the following properties (guaranteed by a number of theorems  - see \cite{Kassel} for the hypotheses and the proofs):
\begin{eqnarray}
&&R R^{op} = \mathbbmss{1}, \qquad (\Sigma \otimes \mathbbmss{1}) R = R^{-1} = (\mathbbmss{1} \otimes \Sigma^{-1})R, \nonumber\\
&&R_{12} R_{13}R_{23} = R_{23} R_{13}R_{12}, \qquad (\Delta \otimes \mathbbmss{1})R = R_{13}R_{23}, \qquad (\mathbbmss{1}\otimes \Delta)R = R_{13} R_{12}. \label{unirel}
\end{eqnarray} 
By representing the relations (\ref{unirel}) in specific representations one recovers the $S$-matrix constraints (resp., braiding unitarity, crossing, Yang-Baxter and bootstrap) which we studied at the beginning of the course \cite{Gustav}. In particular, the antipode has something to do with antiparticle representations, and the coproduct featuring in (\ref{unirel}), as we have earlier motivated, has something to do with the tensor product of two representations (which physically allows to extract the bound state as a particular irreducible component). 

It is important to mention that the universal $R$-matrix has been systematically constructed for all the trigonometric and rational quantum groups based on simple Lie (super)algebras, thanks to a series of work by the mathematicians S. Khoroshkin and V. Tolstoy.

Versions of the $R$-matrix (\ref{erre}) appear in the study of the RTT relations of the XXZ spin-chain (see for instance section 10 in \cite{FaddeevHow}). Reference \cite{FaddeevHow} displays a very instructive example of RTT relations and the method of the algebraic Bethe ansatz for spin-chains. The introduction of non-constant parameters (inhomogeneities on spin-chain sites) can typically be performed via the process of {\it Baxterisation} or by switching to the affine version of the quantum group $U_q\big(\widehat{\mathfrak{sl}}(2)\big)$. For example, one can prove that the $R$-matrix 
\begin{eqnarray}
\label{erreb}
R(\theta)=\begin{pmatrix}\sinh (\eta+ \theta)&0&0&0\\0&\sinh \theta&e^\theta \, \sinh \eta&0\\0&e^{-\theta} \, \sinh \eta&\sinh \theta&0\\0&0&0&\sinh (\eta+\theta)  \end{pmatrix},
\end{eqnarray}
dependent on the additional continuous parameter $\theta$ for any fixed $\eta \in \mathbbmss{C}$,
satisfies (\ref{quasico}) with respect to a version of the $U_q\big(\mathfrak{sl}(2)\big)$ quantum group symmetry given by
\begin{eqnarray}
\label{copo}
\Delta(E) = E \otimes \mathbbmss{1} + q^{-H} \otimes E, \qquad \Delta(F) = F \otimes q^H + \mathbbmss{1} \otimes F, \qquad \Delta(H) = H \otimes \mathbbmss{1} + \mathbbmss{1} \otimes H,
\end{eqnarray}
in the representation (\ref{repa}), and having set
\begin{eqnarray}
q = e^\eta.
\end{eqnarray}
The parameter $q$ ultimately depends on the anisotropy parameter $\Delta$ in the XXZ Hamiltonian, but not on the exchange coupling $J$. One can find in the literature that the anisotropy parameter is connected with $\frac{q+q^{-1}}{2}$. As $q \to 1$ one obtains the Heisenberg spin-chain, displaying an undeformed $\mathfrak{su}(2)$ symmetry (which also extends to the ``Baxterised'' version, namely the Yangian of $\mathfrak{su}(2)$ \cite{Loebbert}). In the Heisenberg spin-chain, the sign of $J$ determines whether the ground state is ferromagnetic, namely all spins aligned, or antiferromagnetic, namely maximum disalignment.

{\it Exercise [3 hours' work]: verify that also $\ref{copo}$ is legitimate coproduct for $U_q\big(\mathfrak{sl}(2)\big)$, in the sense that it respects the defining relations. Then verify that (\ref{erreb}) is an $R$-matrix for this coproduct.}

{\it Exercise [1 hours' work]: write a Mathematica programme which verifies that (\ref{erreb}) satisfies the Yang-Baxter equation. For this it will be useful to write $\theta = \theta_1 - \theta_2$, such that $R_{13}$ will be a function of $\theta_1 - \theta_3$ {\it etc.}} 

The $R$-matrix (\ref{erreb}) directly controls the RTT relations of the XXZ spin-chain - see for instance \cite{Anastasia} formula (4.11). This $R$-matrix is in the same class as the $R$-matrix of the Sine-Gordon soliton-antisoliton scattering (see also chapter 18 in \cite{Mussardo}), belonging to the family of trigonometric quantum-group $R$-matrices \cite{AleDurham}. The association of the Sine-Gordon model with the quantum group $U_q\big(\widehat{\mathfrak{sl}}(2)\big)$ is well known and can in fact be derived via purely field-theoretic methods  - see \cite{BernardLeClair} (particularly their formulas (3.39a) to (3.39c) and their appendix B) and \cite{Reshe}. The parameter $q$ is related\footnote{Notice that this connection bears a certain degree of ambiguity, and different sources present different deformation parameters $q$ with slight variations in the coupling-constant dependence. It is however clear that a relationship of the type (\ref{inhereq}) will hold.} to the dimensionless coupling $\beta$ by
\begin{eqnarray}
\label{inhereq}
q = \exp \Big[\frac{16 \pi^2 i}{\beta^2}\Big].
\end{eqnarray}  
Notice that at the ``free-fermion'' point $\beta^2 = 4\pi$ and at the marginal point $\beta^2 = 8 \pi$ we have $q \to 1$, recovering thereby an undeformed $\mathfrak{su}(2)$ symmetry as a subalgebra of the total symmetry of the model.  At the special supersymmetric value $\beta^2  = \frac{16 \pi}{3}$ we see that $q=-1$. We now need to recall the the eigenvalues of $H$ are $\pm \frac{1}{2}$, to conclude that effectively the coproduct develops a statistical nature (a hint, in fact a necessary but not sufficient condition for an underlying supersymmetry of the theory) but with fractional fermion number $\pm \frac{1}{2}$. The existence of fractional fermion numbers is therefore intimately connected with the quantum group (non-local) nature of the symmetry algebra. Notice that this is not immediately related to the statistics of the sine-Gordon soliton/antisoliton, since there is an issue of local vs. non-local realisation of the symmetry \cite{iFendley}. The local realisation of the soliton/antisoliton fermion number is the topological charge which is additive ({\it i.e.} it acts via a standard coproduct, in Hopf algebra parlance).

{\it Project [5 weeks' work]: explore the literature on fractional fermion numbers in $1+1$ dimensions, taking the moves from \cite{iFendley} but also from \cite{Roman}. Revisit and expand the considerations we have made in this section in the light of such literature review.} 

\section{The Thirring model}

The Lagrangian of the massive Thirring model is 
\begin{eqnarray}
L_T = \bar{\psi} (i \gamma^\mu \partial_\mu - m_T)\psi - \frac{g}{2} \bar{\psi} \gamma^\mu \psi \, \bar{\psi}\gamma_\mu \psi, 
\end{eqnarray}
in terms of a Dirac fermion field in $1+1$ dimensions, with the associated Clifford-algebra gamma matrices \cite{Tong}:
\begin{eqnarray}
\gamma^0 = \sigma_1, \qquad \gamma^1 = i \sigma_2 \qquad \longrightarrow \qquad  \gamma^3 = \gamma^0 \gamma^1 = -\sigma^3.
\end{eqnarray}
The fermion field has engineering mass-dimension $\frac{1}{2}$ so the four-fermion term is marginal in $1+1$ dimensions, with $g$ a-dimensional. The model is known to be quantum-mechanically well-defined for $g> -\pi$ \cite{Coleman}. The limit $g\to-\pi^+$ is the limit of {\it strong repulsion} \cite{KorepinRepu}: the name will make a lot of sense from the point of view of the duality with Sine-Gordon - see eq. (\ref{relziu}), which implies $\beta^2 \to \infty$ in this limit. For $g>-\pi$ the theory is renormalisable - see for instance \cite{Luther,Luscher}. The coupling $g$ is not renormalised, but the mass is.

The massive theory is integrable \cite{Mik}, and $\mathfrak{u}(1)$-invariant under the global phase shift
\begin{eqnarray}
\psi \to e^{i \delta} \psi, \qquad \delta \in \mathbbmss{R}
\end{eqnarray}
(which we could call ``electric''). 
The associated conserved charge will be dual to the topological charge of the Sine-Gordon solitons/antisolitons. In fact under the duality conjectured by Coleman the elementary Thirring fermion should behave like the soliton in Sine-Gordon (see the next section). Quantum integrability follows partly from the duality with Sine-Gordon, but more intrinsically from the relationship with the XXZ spin-chain (see one of the exercises in the next section). As in the case of the Sine-Gordon model, the XXZ chain can be thought of as a lattice regularisation of the Thirring model \cite{Luther}.

Setting $m_T=0$ gives the massless Thirring model, which is also invariant under chiral transformations - rotating independently the right and left spinor components $\frac{1\pm\gamma_3}{2}  \psi$. In addition, the massless model is scale-invariant. 

\section{\label{Equivalence}Duality between Sine-Gordon and Thirring}

We now come to the core of the duality, of which we have already disseminated a few hints. It was stated most effectively by Coleman \cite{Coleman} and generated a massive amount of activities, highlighting some of the most striking features of (integrable) two-dimensional quantum field theories. 

The statement which Coleman argued was as follows: there is a relationship between the two theories, established by 
\begin{enumerate}
\item a relation between the parameters:
\begin{eqnarray}
\frac{4\pi}{\beta^2} = 1 + \frac{g}{\pi},\qquad \beta^2<8\pi,\label{relziu}
\end{eqnarray}
$\beta^2>0$ being consistent with $g>-\pi$. This is in line with the non-renormalisation of either couplings appearing in (\ref{relziu}). In particular, $\beta^2 = 4 \pi$ corresponds to free fermions $g=0$. It is noteworthy that small $\beta$ corresponds to $g$ going to infinity (an instance of weak-strong duality, akin to S-duality). The relation (\ref{relziu}) is also consistent with the fact that $\beta^2 = 4 \pi$ signals the transition between attractive and repulsive regimes for the soliton (resp. fermion) interaction; 

\item a relation between the fields 
\begin{eqnarray}
-\frac{\beta}{2\pi} \epsilon^{\mu \nu}\partial_\nu \phi = \bar{\psi}\gamma^\mu \psi, \qquad \frac{m^2}{\beta^2} \cos \beta \phi = - Z m_{T} \bar{\psi}\psi,\label{refin}
\end{eqnarray}
to be understood in the sense of perturbative calculations of correlation functions, with $Z$ a constant dependent on the regularisation (intuitively, $Z$ incorporates in fact two different regularisation schemes, one for $m$ and one for $m_T$, but one can bring both ambiguities on one side of the equation). The second formula in (\ref{refin}) is in fact refined to \cite{Coleman,FaberIvanov}
\begin{eqnarray}
-Z m_{T} \bar{\psi} \Big(\frac{1\mp \gamma_3}{2}\Big)\psi = \frac{m^2}{\beta^2} e^{\pm i \beta \phi}.\label{that}
\end{eqnarray}

Note that Coleman used a very specific perturbation theory\footnote{Effectively using what is known as Conformal Perturbation Theory.}, performed around vanishing mass, with a delicate treatment of the severe infrared issues presented by a massless scalar field in two dimensions. The duality eventually transcends the particular perturbative analysis and regularisation scheme used, and is in fact directly displayed in the exact $S$-matrices.

\end{enumerate}

It is important to note (see also the Conclusions) that the ``dictionary'', as originally stated by Coleman, involves only relations between objects with zero total topological / electric charge. We shall later on see a test which in principle extends beyond this sector (see further on the {\it form factor} test), at least for a particular class of observables.

Perturbative checks at the level of the $S$-matrix are performed for instance in \cite{Weisz} - in the case of Thirring one uses standard (small $g$) perturbation theory (more on this in the next section), in the case of the solitons one can resort to semiclassics \cite{DashenHasslacherNeveu}. These perturbative-type checks of the exact formula are in rather distant regions of validity. In fact the semiclassical regime corresponds to small $\xi$, since, by looking at the $\frac{1}{\hbar} \int d^2x \frac{m^2}{\beta^2} \cos \beta \phi$ term, we see that the dimensionless parameter in Sine-Gordon is $\hbar \beta^2$. Therefore the expansion in $\beta^2$ is also an expansion in $\hbar$. 

It is worth noticing that the Sine-Gordon singularity highlighted by Coleman, occurring as $\beta^2$ reaches $8 \pi$ from below, has a correspondence in Thirring where $g$ reaches $-\frac{\pi}{2}$ from above. Coleman \cite{Coleman} once again provides an argument based on the energy density to demonstrate that it becomes unbounded from below under this critical value. The singularity was later argued to be only apparent \cite{FaberIvanov}.

Although these days, after the advent of AdS/CFT, we are in some sense used to the most dazzling dualities between theories who do not even remotely resemble one another, at the time when Coleman proposed this particular duality he was in fact pointing the finger on a remarkable property of two-dimensional theories. In particular, the rearrangement of some of the degrees of freedom of one theory into those of the other is a highly non-trivial and non-perturbative phenomenon. The soliton is not created by a local field. Mandelstam \cite{Mandelstam} (see also \cite{Marchetti}) has constructed the quantum soliton creation and annihilation operators directly within Sine-Gordon, in terms of normal-ordered exponentials of the (integral of the) boson field and its derivatives, and showed that such operators satisfy the correct anti-commutation relations. The soliton state in fact behaves like a fermion, as we have in a few places anticipated, and in the Thirring description we have an elementary fermionic field. We refer to \cite{Mandelstam, Marchetti} for a discussion of the subtleties associated with identifying the operators in the two theories in the light of the constructive approach to quantum field theory.

The mass relationship depends on the renormalisation scheme used in the two theories, but it is natural to adjust the two schemes such that the finite parts satisfy \cite{LukZam}
\begin{eqnarray}
m_{T,r} = m_{sol},
\end{eqnarray}
where $m_{T,r}$ is the renormalised fermion mass.
The duality is established in the ``zero-charge'' sector (namely, only involving bosonic combinations with zero topological / $\mathfrak{u}(1)$ charge). There exist constructive quantum field theory approaches which can make statements beyond the zero-charge sector \cite{SeilerUhlenbrock}, but they also go beyond the scope of these lectures and of the author's knowledge.

The duality finds a deep realisation within the common integrable structure, ultimately controlled by the underlying quantum group. This is manifested in the Bethe ansatz formulation of the spectral problem, which was studied for instance in \cite{BergknoffThacker}. 
For a review of the method of the Bethe ansatz to solve integrable systems we refer to \cite{FaddeevHow,Anastasia,Fedor,Jules} - see also the next section.

\smallskip

{\it Project [8 weeks' work]: read and reproduce section II of \cite{BergknoffThacker} - and read the rest of that paper. This will show how integrability allows to construct an exact analogue of the procedure of filling the Dirac sea for fermionic quantum field theories, in a fully interacting theory as opposed to free fields (perturbatively). The conservation of the particle number and the use of the Bethe wave-function is crucial to that purpose. An interesting addendum to the project could be reproducing section IV of \cite{BergknoffThacker}: in it an interesting relation is obtained for the renormalised mass, from the viewpoint of the Bethe ansatz. Certain integrals appearing in the continuum limit of the Bethe equations have to be cutoff, and this establishes a connection with the ordinary field-theory renormalisation.}

\smallskip

A fact that we have already remarked upon in the context of Coleman's duality is that the ``free-fermion'' point of the Thirring model is dual to Sine-Gordon at the special value $\beta = \sqrt{4 \pi}$. If we restrict to the massless case, we can understand the duality, even for values of the radius other than $1$, by directly thinking about bosonisation. We recommend the excellent review \cite{Miranda} for a pedagogical treatment of bosonisation - see also \cite{TongNotes} for a discussion of some of the subtleties related to this point.

{\it Literature search [unspecified amount of time]: explore the literature on \underline{massless integrable scattering}, taking the moves from \cite{Fendley:1993jh}, see also \cite{Stefano}. You will discover the issue of left and right movers, and the theory of massless flows. This also should highlight that the massless limit of the quantum Sine-Gordon theory, which in more than one place we have alluded to in rather a cavalier fashion, is indeed quite subtle, as can be seen by carefully analysing its scattering theory - see in particular \cite{Zamolodchikov:1992zr,Clare}. Flesh out the following schematic diagram:

\medskip

\centerline{\begin{tabular}{|l|}
\hline
\\
{\footnotesize \rm massless Sine-Gordon (left,right) scattering and massless TBA $vs.$ free compact boson $vs.$ interacting massless fermion}
\\
\\
\hline
\end{tabular}}

\medskip

\noindent with the second $vs.$ being described by bosonisation \cite{TongNotes}.
The literature on massless integrable scattering is enormous, and  reaches all the way to the most recent applications in holography.} 

{\it Literature search [unspecified amount of time]: It would be interesting to ascertain whether the massless limit of Sine-Gordon bears any memory of the chiral symmetry of massless Thirring, presumably (if at all) in the (left,right) nature of the scattering. Some literature on this might exist, which is beyond this review's author's knowledge.}

\subsection{Bethe Ansatz}

One of the alternative ways of testing these ideas is to regard the Sine-Gordon model and the Thirring model through the lens of the Bethe ansatz \cite{Fedor}, which is controlled by the same quantum group structure and ends up relying on the same representation for both theories. It is not surprising therefore that the Bethe equations one finds when describing the finite-volume spectrum do correlate (and are both in turn connected with the XXZ Bethe ansatz). This gives us an opportunity to explain what the RTT relations are, and to summarise the method of the algebraic Bethe ansatz, as a unified method to treat all the different quantum integrable systems. We shall also see how the complication of the Sine-Gordon spectrum (with respect for instance to the Lieb-Liniger model which we have encountered earlier) forces us towards a {\it nested} structure of Bethe states. 

The Bethe equations can be constructed by employing the tool of the transfer matrix, which is built as the trace of a string of S matrices for an ordered sequence of interacting particles. Let us briefly outline the calculation. Consider $N$ relativistic particles on a circle of length $L$. The natural \underline{quantisation condition} for the momenta can be taken to be (see for example section 3 in \cite{Zamolodchikov:1992zr}):
\begin{eqnarray}
\label{eigen}
e^{-i p_k L} \, \mbox{tr}_0 T(p_k | p_1,...,p_N) |\psi\rangle = |\psi \rangle, \qquad k = 1,...,N,
\end{eqnarray} 
where $p_i$ are the momenta and  the {\it quantum transfer matrix}, namely the trace over the $0$-th space of the {\it quantum monodromy matrix} 
\begin{eqnarray}
\label{traccia}
\Big[T_a^b(p_0|p_1,...,p_N)\Big]_{c_1 ... c_N}^{d_1 ... d_N} = \sum_{\{k\}} S_{a c_1}^{d_1 k_1} (\theta_0 - \theta_1) \,  S_{k_1 c_2}^{d_2 k_2} (\theta_0 - \theta_2)... S_{k_{N-1} c_N}^{d_N b} (\theta_0 - \theta_N),
\end{eqnarray}
appears, $S$ being the $S$-matrix with the indices associated with the internal degrees of freedom completely spelt out (we will often suppress them hereafter). The trace is over the first space, called {\it auxiliary}, all other spaces being called {\it physical} or {\it quantum}. The objects appearing here are now the natural quantum versions of the classical monodromy and transfer matrices, where the path-ordered exponentials is replaced by an ordered product. The $S$-matrix itself (or rather its associated $R$-matrix) is playing the role of a quantum version of the Lax matrix $L$. In Hopf-algebra language, where we have seen it is customary for the indices to denote different spaces in the tensor product, we would write 
\begin{eqnarray}
T_0 = \prod_{i=1}^k R_{0i}.\label{como}
\end{eqnarray}

\begin{figure}
\centerline{\includegraphics[width=10cm]{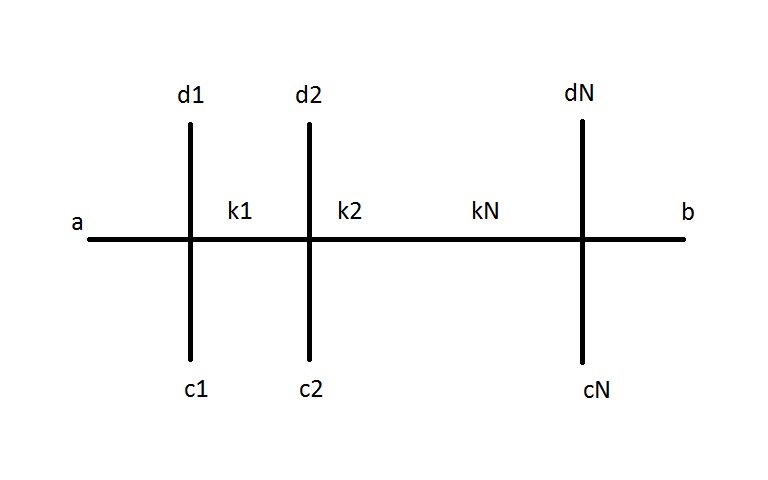}}
\caption{A pictorial representation of the transfer matrix - the auxiliary space being identified as the horizontal line and traced over.}
\label{figura}
\end{figure}

The intuition behind the quantisation condition for the momenta is as follows. When the particle labelled by $k$ goes around the circle of length $L$, it scatters sequentially off all the remaining ones (thanks to the property of factorised scattering). Since the particle has come back to the original position, this must amount to a simple phase shift. We can formalise this by saying that acting on an eigenstate $|\psi \rangle$ of the transfer matrix, and compensating the global phase shift, should result in an identical wave-function. Notice that we should have excluded the same particle $k$ in the sequence of scatterings. We can however formally include it using the fact that $S_{a c}^{d b}(0) = \pm \delta_a^d \delta_c^b$ for the particles which we are interested in. A figure which one often finds being drawn in this context is fig. \ref{figura}.

The task is then to construct the eigenstates $|\psi \rangle$ of the quantum transfer matrix, thereby diagonalising the condition (\ref{eigen}). The algebraic Bethe ansatz achieves precisely this. The crucial insight is that the monodromy matrix satisfies, by sole virtue of the fact that the $S$-matrix satisfies the Yang-Baxter equation, a set of relations called {\it RTT}, named after they appearance:
\begin{eqnarray}
R_{00'} \, (T_0 \otimes \mathbbmss{1}_{0'})\, (\mathbbmss{1}_0 \otimes T_{0'})   = (\mathbbmss{1}_0 \otimes T_{0'}) \, (T_0 \otimes \mathbbmss{1}_{0'}) \, R_{00'},
\end{eqnarray}
where a second auxiliary space $0'$ is introduced.
The quantum spaces are not explicitly indicated but are common to both $T_0$ and $T_{0'}$. The auxiliary variables are $p_0$ and $p_{0'}$, respectively. 

{\it Exercise [45 minutes' work]: Prove that the Yang-Baxter equation for $R$ combined with (\ref{como}) implies the RTT relations.}

Taking the trace $\mbox{tr}_0 \otimes \mbox{tr}_{0'}$ on both sides of the RTT relations shows that trace $\mbox{tr}_0T$ commutes with itself at different values of the auxiliary argument. As in the classical case, by expanding in the auxiliary variable (which here plays the role of a spectral parameter for the quantum problem) one generates an infinite set of charges in involution. If we can prove that one of them is the Hamiltonian\footnote{In ordinary spin-chains the (local) Hamiltonian can be extracted by taking the logarithmic derivative of the transfer matrix evaluated at a special value of the spectral parameter - this is often when all the auxiliary and quantum rapidities (their non-relativistic analogue rather) are set equal \cite{tarasov}. At this point all the $R$-matrices, as we have seen before, tend to degenerate to permutation operators by virtue of the braiding unitarity argument. We thank Julius Julius for conversation on this point.}, these charges are all conserved. To demonstrate how one constructs the eigenvectors we proceed as follows.

We decompose the monodromy matrix as 
\begin{equation}
\label{entri}
T \big(\theta_0|\vec{\theta} \, \big) = E_{11} \otimes A\big(\theta_0|\vec{\theta} \, \big) + E_{12} \otimes B\big(\theta_0|\vec{\theta} \, \big) + E_{21} \otimes C\big(\theta_0|\vec{\theta} \, \big) + E_{22} \otimes D\big(\theta_0|\vec{\theta} \, \big) = \begin{pmatrix}A&B\\C&D\end{pmatrix},
\end{equation}
having singled out the auxiliary space and having conglomerated into a vector all the quantum variables. The trace corresponds to the operator $A+D$. We have used the matrix unities $E_{ij}$, with all zeroes but $1$ in row $i$ and column $j$, in the auxiliary space. The operators $A$, $B$, $C$ and $D$ act on the quantum spaces $1,...,N$. The generic eigenvector is built out of {\it excitations} over a {\it pseudovacuum} state $|0\rangle_p$, and is given by the following expression (for $M$ excitations):
\begin{equation}
|\beta_1,...,\beta_M \rangle = \prod_{n=1}^M B\big(\beta_n|\vec{\theta} \, \big) |0\rangle_p.\label{magn}
\end{equation}
The pseudovacuum is dubbed as such because it is not necessarily the ground state of the theory - in fact in the case of Sine-Gordon it will be the state with $N$ solitons with rapidities $\theta_1,...,\theta_N$. The pseudovacuum has to be a lowest-weigth state of the algebra generated by the $B$ and $C$ operators via the RTT relations. Specifically $|0\rangle_p$ has to be annihilated by $C$ for all value of the parameters.

One can prove, using the RTT relations, that (\ref{magn}) is an eigenvector of the transfer matrix for arbitrary $M$,  thereby diagonalising simultaneously all the commuting charges. Breaking down the RTT relations one obtains formulas which \underline{morally} read as $[A+D,B] \propto B$: this is in all essence the reason why $B$ qualifies as a creation operator.

{\it Exercise [5 hours' work]: Take the $R$-matrix associated with the Sine-Gordon model at $\beta^2 = \frac{16 \pi}{3}$, and write every entry explicitly using the matrix unities $E_{ij}$. Derive all the relations invoving the operators $A,B,C,D$ which descend from writing the RTT relations in components. You will not find ordinary commutation relations, rather what are known as ``exchange relations'', defining an algebra and not a Lie algebra. The exchange algebra is actually another way of writing the quantum group controlling the symmetry of the system (the so-called ``RTT presentation'' of the quantum group). You can use \cite{FaddeevHow} as a guide - see for instance the formulas (66)-(68) in \cite{FaddeevHow}, which are written for the $R$-matrix of the Heisenberg spin-chain. After you have done the first couple of components you are welcome to write a Mathematica code which generates them all}. 
 
The nesting for Sine-Gordon comes at this stage. It is actually not enough to act with the $B$ operators on the pseudovacuum to create an eigenstate, since there is an extra ({\it unwanted}) term which one needs to cancel. This can be achieved by imposing a first set of Bethe equations, algebraic conditions linking the excitation rapidities $\beta_m$ to the rapidities $\theta_i$ of the initial (sometimes called {\it frame}) particles. Such equations may be referred to as {\it level-one} Bethe equations. Only after imposing them, another set of {\it momentum-carrying} Bethe equations finally follow from (\ref{eigen}), linking all the rapidities (frame and excitations) to the circle-length $L$ via the transfer matrix-eigenvalues. The notion is that the momentum is assigned to each particle, but one can still excite the internal degrees of freedom - by acting with $B$ one does this very much in the spirit of using the root generators of a Lie algebra to ``create'' the various states in a symmetry multiplet.

The energy is one coefficient in the expansion of the transfer matrix eigenvalue, and for Sine-Gordon it will read
\begin{eqnarray}
E_{tot}=\sum_{i=1}^N m_{sol} \, \cosh \tilde{\theta}_i,\label{totala}
\end{eqnarray}
with $\{\tilde{\theta}_i\}_{i=1,...,N}$ being each of the possible sets of simultaneous roots of the Bethe equations (quantum configurations). Notice that finding the possible sets of simultaneous Bethe roots $\{\tilde{\theta}_i\}_{i=1,...,N}$ involves finding at the same time the corresponding sets of auxiliary roots $\{\tilde{\beta}_j\}_{j=1,...,M}$, solutions to the auxiliary Bethe equations. Integrability is again at work, allowing the Bethe ansatz to produce an additive free-particle-type formula (\ref{totala}) for the energy (the dynamics being stored in the $\tilde{\theta}_i$ and in $m_{sol}$). Finding the (auxiliary and momentum-carrying) Bethe roots for arbitrary $N,M,L$ is a formidable problem\footnote{We acknowledge discussion with Bogdan Stefa\'nski and Yang-Hui He on this. There is an enormous literature on solutions to the Bethe equations, in fact one might say that this is what integrability ultimately leads to. For the most recent work which we can think  of please see \cite{Rafael}.}. 

The explicit set of Bethe equations for Sine-Gordon (including the breathers, which we have ignored so far) can be found for instance in \cite{Franzini}. As we mentioned earlier, the pseudovacuum is typically made of $N$ solitons, and the excitations correspond to ``flipping'' $M$ of them into antisolitons. In the domain of spin-chains these kinds of excitations, obtained by flipping internal degrees of freedom (such as spin for the XXZ chain), are called often {\it magnons}. In Sine-Gordon each $B$ operator creates an antisoliton from a soliton. One needs a different set of operators to create the solitons from the true vacuum in the first place - this extra ``nested'' step is missing in the Lieb-Liniger model for example, where one can construct immediately $B$ operators which create the only particles in the game out of the perturbative vacuum.   

The Thirring model Bethe ansatz \cite{BergknoffThacker}, object of the {\it Project} outlined in the previous section and eventually correlating with the Sine-Gordon Bethe ansatz \cite{KoreBr}, can be seen to be performed using a variant of the algebraic Bethe ansatz, called the {\it coordinate} Bethe ansatz (which historically predates the former). As we have had the opportunity of highlighting with the {\it Project} laid out in the previous section, the full detail of the Bethe ansatz construction suited for the duality is highly non-trivial. We refer again to \cite{KorepinRepu,BergknoffThacker,KoreBr} for the complete descripton. 

\subsection{Form-Factor Test}

Another one of the tests of the conjecture is to compute form factors, using the exact Sine-Gordon $S$-matrix, for a selection of operators which are easily identifiable on either side of the duality, and compare with explicit Feynman-diagram computations. 

The $n$-particle form factor associated with an operator $\cal{O}$ is defined as
\begin{equation}
\label{in}
F^{\cal{O}}_{\alpha_1 ... \alpha_n} (\theta_1,...,\theta_n) = \langle 0| {\cal{O}}(0)|\theta_1,...,\theta_n\rangle_{\alpha_1 ... \alpha_n},
\end{equation}
where $\theta_i$ is the rapidity of the $i$-th particle {\it in} state, and $\alpha_i$ collectively represents any polarization the particle has - effectively, the Cartan-subalgebra eigenvalues of a suitable (super)algebra representation. 

The axiomatic approach, due in large measure to Smirnov \cite{Smirnov}, to the study of form factors consists of solving a series of rather general conditions, very much like the ones that determine the exact $S$-matrix as we described at the beginning of these lectures. These ``axioms'' too are the historical product of abstracting and axiomatising properties which descend from Feynman diagrams. We present here a summary of these conditions, and refer the reader to \cite{Babu,Weisz,Mussardo,Sergey,BabuF,Diego} for a more complete treatment.

\begin{itemize}

\item {\it Permutation axiom (or Watson's equation)}
\begin{eqnarray}
&&F^{\cal{O}}_{\alpha_1 ... \alpha_{j-1} \, \beta_{j} \, \beta_{j+1} \, \alpha_{j+2} ... \alpha_n} (\theta_1,...,\theta_{j-1} , \theta_{j} , \theta_{j+1} , \theta_{j+2}, ... \theta_n) = \\
&&\qquad \qquad F^{\cal{O}}_{\alpha_1 ... \alpha_{j-1} \, \alpha_j \, \alpha_{j+1} \, \alpha_{j+2} ... \alpha_n} (\theta_1,...,\theta_{j-1} , \theta_{j+1} , \theta_j, \theta_{j+2}, ... \theta_n) \, S^{\alpha_j \alpha_{j+1}}_{\beta_j \beta_{j+1}} (\theta_j - \theta_{j+1}),\nonumber
\end{eqnarray}
where the entries of the $S$-matrix are extracted as
\begin{equation}
S : V_1 \otimes V_2 \longrightarrow V_2 \otimes V_1, \qquad S |v_\alpha (\theta_1) \rangle \otimes |v_\beta (\theta_2)\rangle = S^{\rho \sigma}_{\alpha \beta} (\theta_1 - \theta_2) |v_\rho (\theta_2) \rangle \otimes |v_\sigma (\theta_1)\rangle.
\end{equation}

\item {\it Periodicity}
\begin{eqnarray}
&&F^{\cal{O}}_{\alpha_1 \, \alpha_{2} ... \alpha_{n-1} \, \alpha_n} (\theta_1 + 2i \pi,\theta_{2}, ... \theta_{n-1}, \theta_n) = (-)^\sigma\, F^{\cal{O}}_{\alpha_2 \, \alpha_{3} ... \alpha_{n} \, \alpha_1} (\theta_2 , \theta_{3}, ... \theta_{n} , \theta_1),
\end{eqnarray}
where $\sigma$ is a statistical factor which graphically is often associated with somehow permuting the particle $1$ through the operator (see the pictorial representations in section 7 of \cite{Diego}). In formulas, $(-)^\sigma = (-)^{deg[1] \times deg[{\cal{O}}]}$, where $deg$ is the fermionic degree of an object ($0$ for bosons, $1$ for fermions\footnote{Since we are in $1+1$ dimensions, all the the formulas in this section can be extended to accomodate even more general statistics.}). 

\item {\it Lorentz boost}
\begin{eqnarray}
F^{\cal{O}}_{\alpha_1 \, \alpha_{2} ... \alpha_{n-1} \, \alpha_n} (\theta_1 + \Lambda,\theta_{2}+\Lambda, ... \theta_{n-1}+\Lambda, \theta_n+\Lambda) = e^{\mbox{s} \Lambda}F^{\cal{O}}_{\alpha_1 \, \alpha_{2} ... \alpha_{n-1} \, \alpha_n} (\theta_1,\theta_{2}, ... \theta_{n-1}, \theta_n),\label{spino} 
\end{eqnarray}
where $\mbox{s}$ denotes the Lorentz spin of $\cal{O}$ \cite{Patrick}. 

\item {\it Kinematical singularities}

The form factors are meromorphic functions of the rapidities and have a number of kinematical poles, satisfying
\begin{eqnarray}
\label{reso}
&&-\frac{i}{2} \, \mbox{Res}_{\theta_1 = \theta_2+i\pi} F^{\cal{O}}_{\bar{\alpha}_2 \, \alpha_{2} ... \alpha_{n-1} \, \alpha_n} (\theta_1,\theta_{2}, ... \theta_{n-1}, \theta_n) = \\
&&\qquad \qquad {\bf C}_{\bar{\alpha}_2\beta_2} \, \Big[\mathbbmss{1} - (-)^\sigma S_{\alpha_n \rho_{n-3}}^{\beta_n \beta_2}(\theta_2 - \theta_n)...S^{\beta_3\rho_1}_{\alpha_3\alpha_2}(\theta_2-\theta_3)\Bigg]F^{\cal{O}}_{\beta_3 \, \beta_{4} ... \beta_{n-1} \, \beta_n} (\theta_3,... \theta_n), \nonumber
\end{eqnarray}
where $\bar{\alpha}$ indicates the antiparticle of $\alpha$ and ${\bf C}_{\bar{\alpha} \beta}$ represents the charge-conjugation matrix (in Sine-Gordon, this will be ${\bf C}_{\bar{\alpha} \beta} = \delta_{\alpha\beta}$, $\alpha,\beta = s,\bar{s}$, with the understanding that $\bar{\bar{s}}=s$). The statistical factor $(-)^\sigma$ is obtained by graphically permuting the left-most state through $\cal{O}$: namely, $(-)^\sigma = (-)^{deg[2] \times deg[{\cal{O}}]}$ (we refer again to section 7 in \cite{Diego}, combined with the fact that the fermionic degree of an antiparticle is the same as the particle itself). By $\mathbbmss{1}$ in (\ref{reso}) one means $s_{\alpha_n \rho_{n-3}}^{\beta_n \beta_2}...s^{\beta_3\rho_1}_{\alpha_3\alpha_2}$, with $s^{ab}_{cd} = \delta^a_c\delta^b_d$.

It is quite fiddly to locate the correct indices, so here is the schematic contraction of $S$-matrices in the kinematical singularity condition, for $n=6$:
\begin{eqnarray}
{\bf C}_{\bar{\alpha}_2\beta_2} S^{\beta_6 \beta_2}_{\alpha_6\rho_3}S^{\beta_5 \rho_3}_{\alpha_5\rho_2}S^{\beta_4 \rho_2}_{\alpha_4\rho_1}S^{\beta_3 \rho_1}_{\alpha_3\alpha_2}F_{\beta_3\beta_4\beta_5\beta_6}.  
\end{eqnarray}
\item {\it Bound state singularities}

If the spectrum contains bound states made out of the particles present in the {\it in} state, the form factors will reflect this by having additional poles. These poles are located wherever a pair of rapidities can form a bound state. The residue at these poles has to be proportional to the form factor having as the {\it in} state the bound state and the remaining $n-2$ particle (if the first two particles form a bound state for instance, then this means $F^{\cal{O}}_{{\bf B},a_3,...,a_n}$, where ${\bf B}$ is the bound state of $1$ and $2$). If particles $1$ and $3$ form a bound state, we can use the permutation formula first to bring them close to one another. For the complete description we refer for example to \cite{Diego}. This bound-state condition effectively reduces the calculation to a form factor with one less particle, the same way as the kinematical-singularities condition reduced to two less particles. Often such conditions are powerful enough to generate recursions which produce all the form factors starting from a few initial ones \cite{Mussardo}. Notice that some form factors might be zero due to conservation laws or superselection rules (for instance, form factors must be globally bosonic objects). 
\end{itemize}

The form factor programme only relies on basic  requisites such as the spin of the operator. This is also a slight disadvantage, since one does not begin by knowing the operator, and has to identify it subsequently. The spirit of the form-factor programme, very much like the $S$-matrix programme, consists in by-passing perturbation theory and simply abstracting the experience from Feynman graphs into general principles. The ``axioms'' in fact derive in large part from structural requirements of local relativistic quantum field theory (for an expression of their ``raw'' derivation from first postulates the reader can refer once again to \cite{Diego}). Integrability allows to further massage such postulates into a set of functional relations which we can hope to actually solve.

Correlation functions are obtained by summing over form factors, after suitably inserting resolutions of the identity expanded over multi-particle states. For instance, the two point function is calculated as
\begin{eqnarray}
&&\langle 0| {\cal{O}}(i R,0) {\cal{O}}(0,0) |0\rangle \nonumber\\
&&\qquad =  \sum_{n=1}^\infty \frac{1}{n!} \int_{-\infty}^\infty \frac{d \theta_1}{2 \pi} ... \int_{-\infty}^\infty \frac{d \theta_n}{2 \pi} \sum_{\{\alpha_i\}}F^{\cal{O}\,*}_{\alpha_1,...,\alpha_n}(\theta_1,...,\theta_n) F^{\cal{O}}_{\alpha_1,...,\alpha_n}(\theta_1,...,\theta_n) e^{- R \sum_{i=1}^n m_i \, \cos \theta_i},\nonumber
\end{eqnarray}
$m_i$ being the particle masses, and $|0\rangle$ being the true vacuum of the theory\footnote{Technically one should speak of Wightman functions and equal-time correlators. We shall refer to the literature for more rigorous definitions and specifications, and admit our lack of sufficient knowledge in axiomatic quantum field theory. We thank the LonTI participants for very stimulating conversations on this point.}. This is thanks to the fact that in integrable system we have control on the resolution of the identity via multiparticle states \cite{Diego}:
\begin{eqnarray}
\mathbbmss{1} = \sum_{n=1}^\infty \frac{1}{n!}\int_{-\infty}^\infty \frac{d \theta_1}{2 \pi} ... \int_{-\infty}^\infty \frac{d \theta_n}{2 \pi} \sum_{\{\alpha_i\}} |\theta_1,...,\theta_n\rangle_{\{\alpha_i\}} \langle \theta_1,...,\theta_n|_{\{\alpha_i\}}.
\end{eqnarray} 

This programme was put to frution in the case of the Sine-Gordon model in \cite{BabuF} and a series of related works, using the so-called {\it off-shell Bethe ansatz} method. We shall not discuss alternative techniques, such as for example the method put forward by Lukyanov and Zamolodchikov \cite{LK}. The main result of \cite{BabuF} is summarised as follows.

The form factor $F_{\alpha_1,...,\alpha_n}(\theta_1,...,\theta_n)$ for an {\it in} state formed of solitons and antisolitons is given by the component $(\alpha_1,...,\alpha_n)$ of the row vector
\begin{eqnarray}
N_n^{\cal{O}} \int_ {C_{\underline{\theta}}} d_{u_1} ...\int_{C_{\underline{\theta}}} d_{u_m} \, g(\underline{\theta},\underline{u}) \, \Omega_n \, C_{1...n}(\underline{\theta},u_1)...C_{1...n}(\underline{\theta},u_m),
\end{eqnarray}
where the $C$ operators are the $C$ entries of the $n$-site monodromy matrix built with the Sine-Gordon soliton-antisoliton $S$-matrix (more precisely, the associated $R$-matrix), acting from the right on the pseudovacuum {\it covector}, which one sets to be $\Omega_n = \langle s| \otimes ... \langle s| $ - this will produce as a result a row vector instead of a column vector. The number $m$ counts how many off-shell excitations are created over the pseudovacuum. That is to say, $m$ counts the number of `off-shell magnons' - off-shell waves of antisolitons moving with rapidity $u_j$ in a sea of solitons. All of the solitons of the sea are appearing with their associated inhomogeneity/momentum-carrying rapidity $\theta_i$. Notice that the choice of the pseudovacuum is independent on the fact that the correlation function one later constructs is purported to be on the true vacuum of the theory - the pseudovacuum being here a mere trick to solve Watson's equation. One has introduced the collective notations
\begin{eqnarray}
\underline{\theta} = (\theta_1,...,\theta_n), \qquad \underline{u} = (u_1,...,u_m).
\end{eqnarray}
This method is called of the off-shell algebraic Bethe ansatz, because the auxiliary parameters $u_j$ are not fixed by the Bethe equations (that would be on-shell), but are instead integrated over. The auxiliary roots $\beta_j$ (see previous section) are understood as having been solved for, their only role having been to contribute fixing the $\theta_i$ to a true eigenstate of the spectrum.

The function $g$ is constructed for $n>2m$ as
\begin{eqnarray}
g(\underline{\theta},\underline{u}) = \prod_{1\leq i<j\leq n} F(\theta_i-\theta_j) \prod_{i=1}^n \prod_{J=1}^m   \phi(\theta_i-u_J) \prod_{1\leq I<J \leq m} \tau(u_I - u_J) e^{\pm \frac{\mbox{s}}{n-2m} (2 \sum_A u_A - \sum_k \theta_k)},
\end{eqnarray}
$\mbox{s}$ being the spin of the operator, $F$ being the minimal two-particle form-factor block, 
\begin{eqnarray}
\phi(u) = \frac{1}{F(u)F(u+i \pi)}, \qquad \tau(u) = \frac{1}{\phi(u)\phi(-u)},
\end{eqnarray}
and the contour $C_{\underline{\theta}}$ being a complicated contour described in \cite{BabuF}. In the case $n\leq 2m$ there should be a change of the pseudovacuum (such as: all antisolitons) which allows this formula to be used. The minimal two-particle form-factor block satisfies the axioms for $n=2$ with the $S$-matrix simply replaced by $S(\theta)$, and reduces to a compact integral representation \cite{KarowskiWeisz}:
\begin{eqnarray}
F(\theta) = -i \sinh \frac{\theta}{2} \times \exp \int_0^\infty \frac{dt}{t}\frac{\sinh \frac{(1-\nu)t}{2}}{\sinh \frac{\nu t}{2} \cosh \frac{t}{2}} \frac{1-\cosh t(1-\frac{\theta}{i\pi})}{2 \sinh t},
\end{eqnarray}
$\nu$ being defined by
\begin{eqnarray}
\nu = \frac{\beta^2}{8 \pi - \beta^2}.
\end{eqnarray}
The minimal form-factor block $F(\theta)$ is not the minimal two-particle form-factor, but it is a useful building block for it (and for arbitrary form factors in fact, since it is used to systematically take care of the functional complication of the dressing phase).
To obtain this formula one can utilise a theorem by Karowski and Weisz \cite{KarowskiWeisz}, which says the following. If the dressing factor can be recast in the form
\begin{equation}
\label{integrorep}
S(\theta) = \exp \int_0^\infty dx \, f(x) \, \sinh \frac{x \theta}{i \pi},
\end{equation} 
then the minimal solution to
\begin{equation}
\label{Watson}
F(\theta) = F(-\theta) \, S(\theta), \qquad F (i \pi - \theta) = F (i \pi + \theta), 
\end{equation}
is given by 
\begin{equation}
\label{eff}
F(\theta) = \exp \int_0^\infty dx \, f(x) \, \frac{\sin^2 \frac{x(i \pi - \theta)}{2 \pi}}{\sinh x}
\end{equation}
(all formulas in this theorem being valid in appropriate regions in the complex rapidity plane). 

The test of the duality is now performed by choosing an operator which is clearly identifiable on both sides of the duality. From what we have learnt, the soliton creation operator on the Sine-Gordon side should behave like the basic fermionic creation operator on the Thirring side. The soliton creation operator is specific enough that we can pinpoint exactly which form factor to choose and how to use the formula we have just decribed. On the other hand, at strong coupling (that is, at weak Thirring coupling) we can simply compute a tree-level Feynman graph involving the fermion field. 

The paper \cite{BabuF} performs such a test (and many more were performed in the literature). At three level, one has in particular
\begin{eqnarray}
\langle 0| \psi(0) |\theta_1 \theta_2 \theta_3\rangle_{\bar{s}ss} = - i g \sinh \frac{\theta_{23}}{2} \times \frac{u(\theta_2) \cosh \frac{\theta_{12}}{2} +u(\theta_3) \cosh \frac{\theta_{13}}{2}}{\cosh \frac{\theta_{12}}{2}\cosh \frac{\theta_{13}}{2}\cosh \frac{\theta_{23}}{2}},
\end{eqnarray}  
where $\alpha_i = s$ is a soliton and $\alpha_i = \bar{s}$ is an antisoliton, $\theta_{ij} = \theta_i - \theta_j$ and $u(\theta) = \sqrt{m_{sol}}\begin{pmatrix}e^{-\frac{\theta}{2}}\\e^{\frac{\theta}{2}}\end{pmatrix}$ is the Dirac polarisation spinor. This expression is then perfectly matched with a strong coupling approximation of the exact Sine-Gordon formula.

It is important to notice that this matching is obtained for an assignment of spin 
\begin{eqnarray}
\mbox{s}=\frac{1}{2}.
\end{eqnarray}
Before the {\it caveat} in the Conclusions will have the time to reproach us, we shall allow ourselves to say that the soliton-antisoliton system is trying to convince us that they are a two-component Dirac particle.

\section{Conclusions}

The duality between the Sine-Gordon and the Thirring model is a remarkable chapter of theoretical physics. In these lectures we have tried to portray it in hopefully accessible terms, leaving out some of the subtleties which the reader is encouraged to independently explore. One of such subtleties is delved into in great detail for instance in \cite{KlassenMelzer}, see also the end of section 1.3 in \cite{Feverati}. We also refer again to \cite{TongNotes}, especially pages 369 and following. Technically, what we have presented should be understood as a duality or a correspondence, and not as an equivalence. The two theories are not related by a (local) change of variables, as the ``dictionary'' which we have discussed does not qualify as such. Rather, certain observables in the zero-charge sectors of either theories obey precise matching patterns. But there are differences in other aspects \cite{KlassenMelzer,Feverati}. We hope that this review will be a sufficient starting point for the reader's individual examination of the field. Our pedagogical choice has been to highlight the commonalities, with the primary purpose of instilling the idea of duality by showing one of its first ground-breaking appearances, and the realisation of the power of integrability in analysing such statements.

We have also not made any attempt at completeness when citing the literature, as that would be an impossible task. This topic was and still is, even after so many years of the community getting to grips with dualities (primarily the AdS/CFT duality \cite{Maldacena}), a striking demostration of the versatility of quantum field theory which could indeed be considered a precurson of many modern developments. To signify the subversive nature of this idea, it is worth repeating one sentence from Coleman's acknowledgments in \cite{Coleman}:

\smallskip

{\it ``[...] Jeffrey Goldstone, Roman
Jackiw, and Andre Neveu for discussion [...], Howard Georgi for guiding me [...], Konrad Osterwalder for
reassuring me of my sanity.'' (from \cite{Coleman}).}   

\smallskip

The thought of these giants reassuring each other of their sanity is certainly one of the most impressive images of the impact which this discovery has had.

\section*{\label{sec:Ackn}Acknowledgments}

The author thanks the organisers of the LonTi programme, and in particular Bogdan Stefa\'nski, Nadav Drukker and Yang-Hui He, for the invitation to give these lectures, for the encouragement and for the very interesting discussions at LIMS; Elli Heyes, Dmitrii Riabchenko, Chawakorn Maneerat and the technical and video support for their kind help during the lectures; and the Royal Institution for hosting the lectures in their inspiring London setting. Many thanks go to all the participants to the lectures, for their enthusiasm, their error-spotting ability and their extremely good questions, all of which have already prompted several revisions of this manuscript. Support from the EPSRC-SFI grant EP/S020888/1 {\it Solving Spins and Strings} is gratefully acknowledged. 

\section*{\label{sec:Data}Data management}

No data beyond those presented in this paper are needed to validate its content.

\end{document}